\documentclass[fleqn,usenatbib]{mnras}

\usepackage{newtxtext,newtxmath}

\usepackage[T1]{fontenc}
\DeclareRobustCommand{\VAN}[3]{#2}
\let\VANthebibliography\thebibliography
\def\thebibliography{\DeclareRobustCommand{\VAN}[3]{##3}\VANthebibliography}

\usepackage{graphicx}	
\usepackage{amsmath}	

\title[X-ray spectral and timing in MAXI J1820+070]{X-ray spectral and timing evolution during the 2018 outburst of MAXI J1820+070}

\author[YaXing Li et al.]{
YaXing Li$^{1,2}$,
Zhen Yan$^{1}$\thanks{E-mail: zyan@shao.ac.cn},
ChenXu Gao$^{1,2}$,
and Wenfei Yu$^{1}$
\\
$^{1}$Shanghai Astronomical Observatory, Chinese Academy of Sciences, 80 Nandan Road, Shanghai 200030, China\\
$^{2}$Chinese Academy of Sciences, University of Chinese Academy of Sciences, 19A Yuquan Road, Beijing 100049, China\\
}

\date{Accepted 2025 February 24; Received 2025 February 17; in original form 2024 July 6}

\pubyear{2024}

\begin{document}
\label{firstpage}
\pagerange{\pageref{firstpage}--\pageref{lastpage}}
\maketitle

\begin{abstract}
We made use high-cadence observations from the $Insight$-HXMT and $NICER$ to scrutinize the spectral and timing evolution during the 2018 outburst of the black hole X-ray binary (BHXRB) MAXI J1820+070.
It's hardness-intensity diagram (HID) displays a ``q''-like track including all the spectral states, along a unique loop in the hard state. The tracks observed in the HID is anticipated in the evolution of the components responsible for Compton and reflection emission. This is substantiated by the relationship between the X-ray luminosity $L_\mathrm{X}$ and photon index $\Gamma$ which exhibits a pattern reminiscent of HID. The distinctive evolution of the reflection component leads to the unique loop in the HID (also in the $L_\mathrm{X}$--$\Gamma$ plane) of hard state.
Additionally, we found a negative correlation between frequency of the type-C quasi-periodic oscillation (QPO) ($\nu_{\mathrm{C,QPO}}$) and the optical depth of the Compton emission ($\tau$), and a positive correlation between $\nu_{\mathrm{C,QPO}}$ and $\Gamma$. These correlations strongly suggest a coupling between the QPO properties and the underlying process responsible for Comptonization. We also found that the last detection of type-C QPO coincide with the transient jet ejection within a timescale of one hour.  
 
\end{abstract}

 



\section{Introduction} 
The majority of Galactic black hole X-ray binaries (BHXRBs)  behave as transient sources that spend years to decades in a quiescent state. Occasionally, they undergo outbursts lasting for weeks to months before returning to quiescence \citep{Yan2015,Corral-Santana2016}. 
These outbursts may involve transitions through distinct accretion states, which are characterized by a variety of X-ray spectral and timing features, along with associated radio emission \citep[see reviews by][]{Remillard2006,Done2007,Belloni2010}. Given their proximity at kiloparsec distances and their dramatic variability on timescales spanning from days to months, BHXRBs serve as excellent targets for real-time monitoring of the accretion flow and jet behavior. High-cadence observations yield valuable insights into the physics of accretion and its interplay with jets across different accretion states.

The outburst usually starts from the hard state (HS) when the luminosity is low. The X-ray spectrum of HS is dominated by Comptonized emission produced from a hot plasma usually called corona \citep{1979Natur.279..506S}, a weak thermal component from the accretion disk is also detected. In this state, the strong low-frequency quasi-periodic oscillations (LFQPOs) and broad line noises (BLNs) are usually detected in the power density spectrum \citep[PDS, e.g.][]{2002ApJ...572..392B,2004A&A...426..587C}.
As the outburst goes on, the disk component becomes stronger. When the spectra are dominated by the thermal component from an optically thick disk extending to the innermost stable circular orbit (ISCO) of the BH \citep{1973blho.conf..343N,1973A&A....24..337S}, the outburst enters the soft state (SS). In contrast to HS, SS shows low X-ray variability amplitude with a few percent fractional root mean square (RMS), and LFQPOs are rarely detected \citep{2012MNRAS.427..595M}. The transition between the HS and SS is called intermediate state (IMS). During the outburst decay, the disk temperature and luminosity decrease. At some point, the corona restrengthens and dominates over the disk component, which indicates the outburst return to HS.

A large fraction of outbursts experience a transition from the HS to SS state (H-S) during the outburst rise, and conversely, from SS to HS (S-H) during the outburst decay \citep[e.g.][]{YuWF2009,Tetarenko2016}. Through the comprehensive X-ray spectral and timing analysis, we can conduct detailed research on different spectral states and state transitions \citep[e.g.][]{Belloni2010}. Notably, the X-ray spectral and timing properties exhibit pronounced changes during these state transitions, providing valuable insights into the changes in innermost accretion flow and/or jet around the BH. 
On one hand, the X-ray timing and spectral properties of the BH are closely correlated. The correlations between the QPO frequency and spectral parameters, such as the photon index, disk temperature, and X-ray flux, have been widely studied. Those correlations have been used to investigate the origin of QPOs, accretion geometries across different spectral states, and similarities and/or differences between various BHXRBs \citep[e.g. ][]{Muno1999, Sobczak2000, Vignarca2003, Debnath2008, Shaposhnikov2009, Motta2011, Stiele2013}. On the other hand, there are many spectral-timing techniques and modeling combine energy spectral and variability information to offer additional perspectives on the accretion and jet during the outburst. For example, phase-lag spectra including lag-frequency spectra and lag-energy spectra are usually used to probe the physical connection of variability in different energy bands \citep[e.g. ][]{Uttley2014, DeMarco2015, WangJY2022}, while RMS spectra are used to study the energy-dependent variability of different signals \citep[e.g. ][]{Stiele2015MNRAS,WangPJ2022, MaX2023}. Additionally, phase-resolved spectroscopy has also been applied to study spectral evolution on the QPO time scale \citep[e.g.][]{Ingram2016, Stevens2016, Nathan2022,gao2023} and frequency-resolved spectroscopy is utilized to explore the X-ray energy distribution across different short time scales \citep[e.g.][]{Gilfanov2000, Axelsson2016}.

\section{MAXI J1820+070}
In this work, we focus on the evolution of the very bright outburst of MAXI J1820+070. The X-ray outburst MAXI J1820+070 was first detected by Monitor of All-sky X-ray Image (MAXI) on March 11, 2018 \citep[MJD 58188,][]{2018ATel11399....1K}. The optical counterpart ASASSN-18ey was discovered by the All-Sky Automated Survey for Supernovae (ASAS-SN) \citep{2014ApJ...788...48S} around several days earlier \citep{2018ApJ...867L...9T, 2018ATel11400....1D}. Soon after, the source was classified as a BHXRB according to the follow-up observations in different wavelengths \citep{2018ATel11418....1B, 2018ATel11420....1B, Uttley2018ATel}. \citet{2020MNRAS.493L..81A} measured the distance of $2.96\pm 0.33$kpc by using the radio parallax method and a jet inclination angle of $63 \pm 3^{\circ } $ with very long baseline interferometry (VLBI) observations. \citet{2020ApJ...893L..37T} estimated the mass of black hole is $8.48_{-0.72}^{+0.79} M_{\odot }$. The spin of the BH was estimated as 0.14 \citep{2021ApJ...916..108Z}.

\begin{figure*}
\centering
\includegraphics[width=1\textwidth]{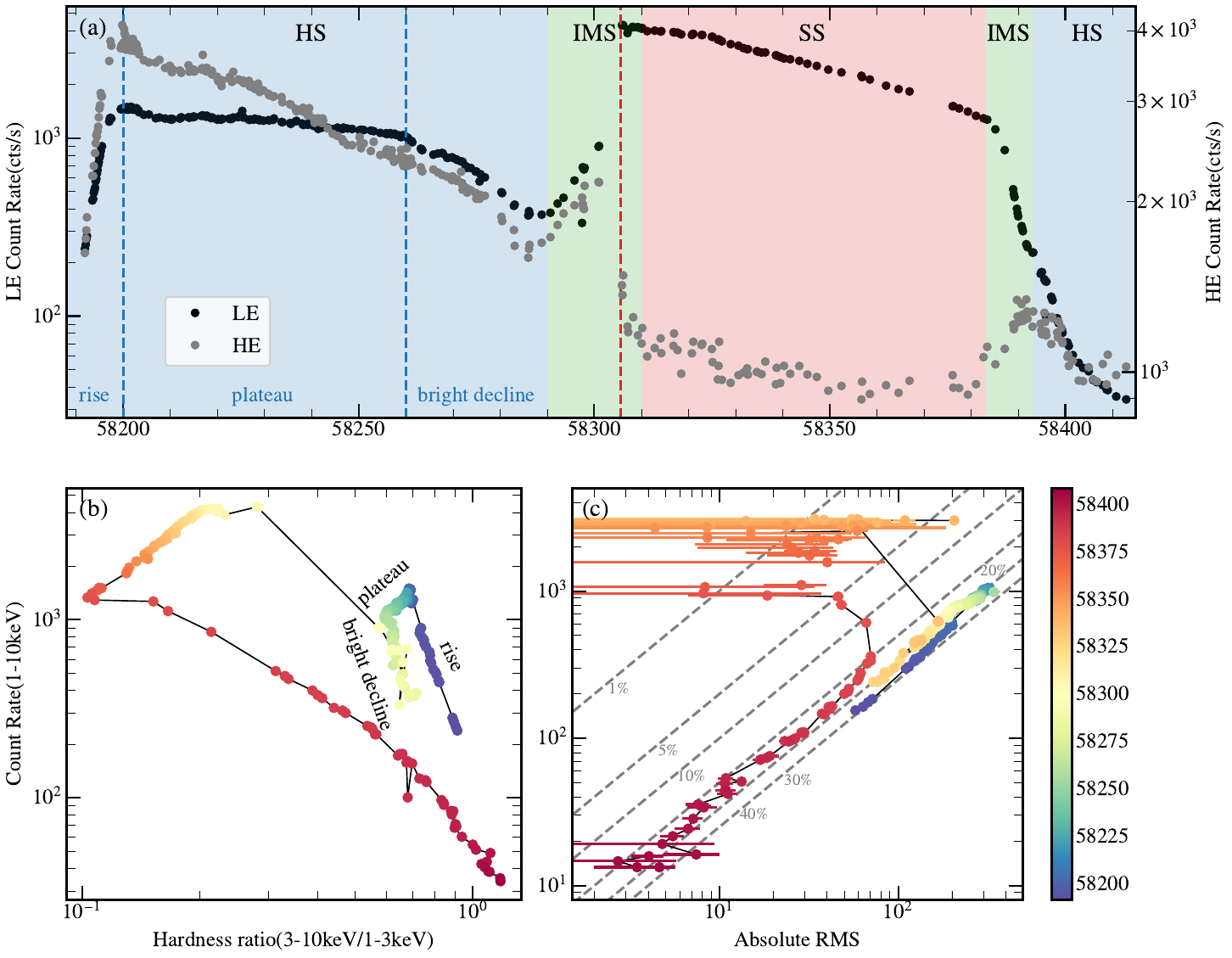}
\caption{$Insight$-HXMT light curve, hardness-intensity diagram and absolute RMS-intensity diagram of MAXI J1820+070 during the 2018 outburst. (a) Light curves obtained from LE and HE of $Insight$-HXMT. The colors of the background represent the spectral states: blue for the hard state, green for the intermediate state and red for the soft state. The blue dashed lines separate three phases of the rising hard state: the rise, plateau and bright decline phases. The red dashed line marks the transient jet ejection time  MJD 58305.60 according to \citet{Wood2021} and also the boundary between HIMS and SIMS.
(b) The $Insight$-HXMT hardness-intensity diagram, defined as the total 1--10 keV count rate vs. the ratio of hard (3--10 keV) to soft (1--3 keV) count rates. The three phases of the rising HS correspond to the three curves of the small ``$\eta$''-like track on the upper right.
(c) The absolute RMS-intensity diagram, with the grey lines corresponding to the values of fractional RMS. Both absolute RMS and fractional RMS are determined in the 0.1--64 Hz from $Insight$-HXMT/LE PDS. The color bar on the right corresponds to the observation time.} 
\label{fig:hidall}
\end{figure*}

Like other transient BHXRBs, MAXI J1820+070 is in HS at the beginning of the outburst. It is a very long HS, extending until around MJD 58290. Subsequently, the source enters the IMS. The above two periods are referred to as rising HS and IMS states. The transient jet ejected around MJD 58306 \citep{Bright2020,Wood2021}, coinciding with the transition between hard intermediate state (HIMS) and soft intermediate state \citep[SIMS; ][]{Fender2004}.
The SS lasts from MJD 58310 to 58383. At about MJD 58393, the source returns to the HS. In order to distinguish from the outburst rise phase, the period between MJD 58383 to 58393 and after are referred to as decaying IMS and HS states, respectively. The time intervals of different spectral states are shown in \autoref{fig:hidall} \citep[see also ][]{2019ApJ...874..183S}. 

The observations targeted on MAXI J1820+070 during this outburst have been conducted by almost all the active X-ray missions in orbit, such as $AstroSat$ \citep{Mudambi2020ApJ}, $Chandra$\citep{Espinasse2020ApJ}, $Insight$-HXMT \citep{YouB2021,Ma2021}, $NICER$\citep{Kara2019,WangJY2021}, $NuSTAR$ \citep{Buisson2019,Buisson2021}, $Swift$ \citep{Stiele2020}, $XMM$-$Newton$ \citep{Xu2020ApJ,Dias2024MNRAS}. Recent studies on the accretion geometry around the black hole in MAXI J1820+070 have made significant strides, yet the conclusions remain a topic of debate. On one hand, the traditional truncated disk/inner hot flow model successfully explains the X-ray spectral and timing properties \citep[e.g.][]{Xu2020ApJ,DeMarco2021,Zdziarski2022ApJ}. On the other hand, a jet-like corona with vertical changes has been proposed to explain the spectral and timing evolution during the hard state \citep[e.g.][]{Kara2019,WangJY2021,YouB2021}. Additionally, complex corona geometries have been suggested, including two corona components producing Compton emission \citep{Buisson2019,Zdziarski2021,MaRC2023} and a stratified hot flow \citep{Dzielak2021MNRAS}.

By concurrently applying X-ray spectral and timing analysis during this outburst, it is very helpful to uncover the properties and evolution of accretion flows around black holes \citep[e.g. ][]{Kara2019, Stiele2020,2020ApJ...896...33W,DeMarco2021,Kawamura2023MNRAS}. 
$Insight$-HXMT is one of the best missions for performing X-ray spectral and timing study in a very broad energy band 1--250 keV \citep{2020SCPMA..6349502Z}. It carries three slat-collimated payloads: low energy X-ray telescope \citep[LE, 1--15 keV,][]{2020SCPMA..6349505C}, medium energy X-ray telescope \citep[ME, 5--30 keV,][]{2020SCPMA..6349504C} and high energy X-ray telescope \citep[HE, 20--250 keV,][]{2020SCPMA..6349503L}. 
All three instruments have large collecting areas and high time resolution, which are beneficial for conducting X-ray spectral and timing analysis for bright targets. $Insight$-HXMT observed MAXI J1820+070 from March 14, 2018 to October 21, 2018, and obtained a total of more than 310 exposures with an average interval of 0.63 days. The $Insight$-HXMT observations extensively cover the different spectral states through almost the entire outburst \autoref{fig:hidall}. We also used the data from $NICER$ in some cases. Due to its high time resolution and large effective area, $NICER$ data can provide X-ray spectral and timing analysis in the 0.2--12 keV.

\section{Data reduction and data analysis}
\subsection{$Insight$-HXMT data reduction}
The $Insight$-HXMT data and products were processed and extracted by using the $Insight$-HXMT Data Analysis Software package (HXMTDAS). We first screened the original event data of three instruments by using \texttt{lescreen}, \texttt{mescreen} and \texttt{hescreen} with the recommended criteria in the user manual \footnote{\url {http://hxmten.ihep.ac.cn/SoftDoc.jhtml}}. The spectra and light curves were extracted by using the tools \texttt{lespecgen}, \texttt{mespecgen}, \texttt{hespecgen}, \texttt{lelcgen}, \texttt{melcgen} and \texttt{helcgen} for the target, and using the tools \texttt{lebkgmap}, \texttt{mebkgmap} and \texttt{hebkgmap} for the background, respectively. All the spectra were grouped with a minimum count of 25 per bin. 

The net light curves were calculated with \texttt{lcmath}. We then used the \texttt{powerspec} in \texttt{XRONOS} package to compute the power density spectrum (PDS) in Leahy normalization \citep{1983ApJ...266..160L}. The net light curves with a time resolution of 1/512 seconds were divided into segments with a length of 512 seconds for producing the PDSs. So the frequency resolution of the PDS is 1/512 Hz and the Nyquist frequency is 256 Hz. 

\subsection{$NICER$ data reduction}
The raw data were first reprocessed by using task \texttt{nicerl2} in the $NICER$ software package \texttt{NICERDAS}. The source and background spectra, ancillary response files (ARFs) and response matrix files (RMFs) were all extracted by using the pipeline task \texttt{nicerl3-spect}. During the above procedure, the SCORPEON background model was chosen, and the recommended systematic error from the calibration file was applied. The spectra were grouped with a minimum count of 25 per bin. 

\subsection{X-ray timing analysis}

We excluded the soft state from MJD 58310 to 58383 for timing analysis of $Insight$-HXMT data. The PDSs were then logarithmically re-binned with an increment factor of 0.025 in frequency and formatted to be compatible with \texttt{XSPEC} \citep{Ingram2012}. We then used a combination of Lorentzian functions and a constant to fit the PDSs in \texttt{XSPEC} (v12.12.1): two or four Lorentzians for the BLNs, two or three Lorentzians for the QPO and its harmonic and/or subharmonic and a constant for the white noise. 
The PDS of the data from LE and ME detectors on the $Insight$-HXMT can be described by a QPO and two BLN components, while the data from HE detectors will use two additional Lorentzians to supply the BLN components. 

Some $NICER$ observations during the IMS were used since the $Insight$ -HXMT observation was absent during the period of MJD 58301--58305 (obsID 1200120192--1200120197). The $NICER$ PDSs consist of one to three BLN components and one or two QPO components. The uncertainties of all parameters were calculated from the Markov Chain Monte Carlo (MCMC) method by the implementation of \texttt{emcee} in \texttt{XSPEC} \footnote{\url{https://github.com/zoghbi-a/xspec_emcee}}. Some examples of PDSs fitting are shown in \autoref{fig:sedpds}. Using the best-fitting results of PDSs, we calculated the fractional RMS $=\sqrt{K/\left \langle x \right \rangle } $, where $K$ is the normalization of the Lorentzian function and $\left \langle x \right \rangle$ is the mean count rate\citep{1983ApJ...266..160L,1989ASIC..262...27V}.

\subsection{X-ray spectral analysis}
\label{3.4}
We performed the spectral fitting through XSPEC v12.12.1 using the \texttt{ chi} statistic. Different models were applied in different spectral states. During the hard state of the outburst rise phase, we applied the model \texttt{constant*TBabs*(thcomp*kerrd+relxillCp+xillverCp)} to fit X-ray spectra in the 2--150 keV. The constant factor was used to coordinate the calibration differences between different instruments, where the constant for LE was fixed at 1. The equivalent hydrogen column density ($N_H$) of \texttt{TBabs} was fixed at $0.15 \times 10^{22}$ cm$^{-2}$ \citep{Uttley2018ATel}. 
The \texttt{kerrd} is a standard optically thick accretion disk model for a Kerr black hole \citep{Ebisawa2003}, which accounts for the accretion disk radiation. We adopted \texttt{thcomp} to describe the Comptonization emission from the corona, where the seed photons come from the accretion disk. And \texttt{ relxillCp} (version 2.3) is a relativistic reflection model that calculates the reflection spectrum from an accretion disk illuminated by the corona \citep{2014ApJ...782...76G,2022MNRAS.514.3965D}, while \texttt{xillverCp} was used to account for a further away and non-relativistic reflection component \citep[e.g. ][]{Buisson2019,YouB2021}.
In our spectral fitting, we fixed the black hole spin at a = 0.14 \citep{2021ApJ...916..108Z}, the inclination angle at $i=63^{\circ } $\citep{2020MNRAS.493L..81A}, the distance at 2.96 kpc \citep{2020MNRAS.493L..81A} and the black hole mass at 8.48$M_{\odot }$ \citep{2020ApJ...893L..37T}. Additionally, Tcol/Teff in \texttt{kerrd} was fixed at 1.7 \citep{Shimura1995ApJ}. We tied the inner disk radius $R_\mathrm{in}$ (in units of $R_{g}$) in \texttt{relxillCp} to that in \texttt{kerrd}. 
For simplicity, we used a continuous power-law emissivity and fixed the index at $q_{1} = q_{2} = 3.0$, and a constant ionization (\texttt{iongrad\_type=0}) in \texttt{relxillCp}. In particular, the reflection fraction \texttt{refl\_frac} in \texttt{relxillCp} and \texttt{xillverCp} were set at -1 to acquire the reflection component. 

Furthermore, we used \texttt{cflux} to get the X-ray flux of different spectral components in the 0.1--100 keV range. Notably, the X-ray flux of the Comptonization component can be calculated as $F_\mathrm{C} = F_\mathrm{thcomp*kerrd} - (1-$\texttt{cov\_frac}$)*F_\mathrm{D}$, where $F_\mathrm{thcomp*kerrd}$ is the integrated flux of the convolution between
\texttt{thcomp} and \texttt{kerrd} components. The total reflection flux is the sum of relativistic reflection flux and non-relativistic flux. However, the non-relativistic component in the IMS is weak and cannot be well constrained; therefore, we only used the \texttt{relxillCp} during this period. During and after the soft state, we applied the model \texttt{constant*TBabs*thcomp*kerrd} to fit the broadband X-ray spectrum. The electron temperature $kT_{e}$ is fixed at 100.0 keV. For some observation during the decaying HS, the covering fraction \texttt{cov\_frac} is fixed at 1 (see \autoref{fig:sedpara}). Then, we also used \texttt{cflux} to calculate the flux of the disk and Comptonization component in 0.1--100 keV energy range. The uncertainties of all the parameters were calculated from the MCMC method using the \texttt{emcee} in \texttt{XSPEC}. All the spectra can be well fitted by our models across different spectral states. We show some examples of energy spectra fitting in \autoref{fig:sedpds}.

Because of the absence of $Insight$-HXMT observation around MJD 58301 to 58305 in the IMS, we used the $NICER$ data as the supplementary. Then we applied the same model as used for $Insight$-HXMT data during this period to fit the spectra in the 1.5--10 keV: \texttt{constant*TBabs*(thcomp*kerrd+relxillCp)}. For simplicity, we fixed the electron temperature $kT_{e}$ at 100.0 keV for those $NICER$ observations. We checked the simultaneous $NICER$ and $Insight$-HXMT  observations before and after this period. The main spectral parameters are almost similar during the HIMS  (see \autoref{app:appb} in detail).

\section{X-ray spectral and timing evolution during the 2018 outburst}

\subsection{Unique evolution of rising HS}

The count rates of MAXI J1820+070 from LE and HE of $Insight$-HXMT are plotted in \autoref{fig:hidall}a, and the hardness-intensity diagram (HID) and absolute RMS-intensity diagram (RID) are also shown in \autoref{fig:hidall}b and c. The light curve of MAXI J1820+070 shows a multi-peaked outburst profile \citep{ChenW1997}. The first peak remains in the hard state, while its HID exhibits a small loop, which looks like a ``$\eta$'' track (\autoref{fig:hidall}b).  
The rising HS is divided into three phases: the rise, plateau and bright decline phases \citep[e.g.][]{Stiele2020, DeMarco2021}, which roughly corresponds to the three curves of the small ``$\eta$''. The rise phase is from the start of the outburst to MJD 58200 when the count rate reaches the peak and then begins to decrease slowly. The plateau phase is from MJD 58200 to 58260, and the bright decline phase is from 58260 to MJD 58290. The energy spectra and PDSs in different phases are shown in \autoref{fig:sedpds}.
The absolute RMS and count rate do not follow a single positive correlation \citep[\autoref{fig:hidall}c; see also][]{Stiele2020}. This loop evolution of rising HS in HID/RID has never been shown in other BHXRBs \citep{Belloni2010,Belloni2016}. Even in other outbursts with multiple peaks, their HID/RID usually do not show a loop trend during the rising HS, such as the outbursts in XTE J1550$-$564, GX339-4 and GRO J1655$-$40 \citep[e.g.][]{Sobczak2000,Debnath2008,Clavel2016}. 

During the rising HS, our timing and spectral analysis results show a clear distinction between this special plateau phase and the rise and bright decline phases. 
Notably, despite the gradual decline of the HE count rate during the plateau phase, the Comptonization luminosity $L\mathrm{_{C}}$ remains almost unchanged. The mass accretion rate $\dot{M}$ and disk luminosity $L\mathrm{_{D}}$ also remain stable during the plateau phase. The evolution trends observed in $\Gamma$, $\dot{M}$ and \texttt{cov\_frac} differ between these three phases of HS. The QPO frequency $\nu_\mathrm{C,QPO}$ exhibits a positive and parallel correlation with the total X-ray luminosity $L_\mathrm{X}$ during the rise and bright decline phases while the correlation is negative in the plateau phase (see \autoref{fig:qpoLGTao}a).

\subsection{X-ray luminosity and spectral parameter evolution}

\begin{figure*}
\centering
\includegraphics[width=1\textwidth]{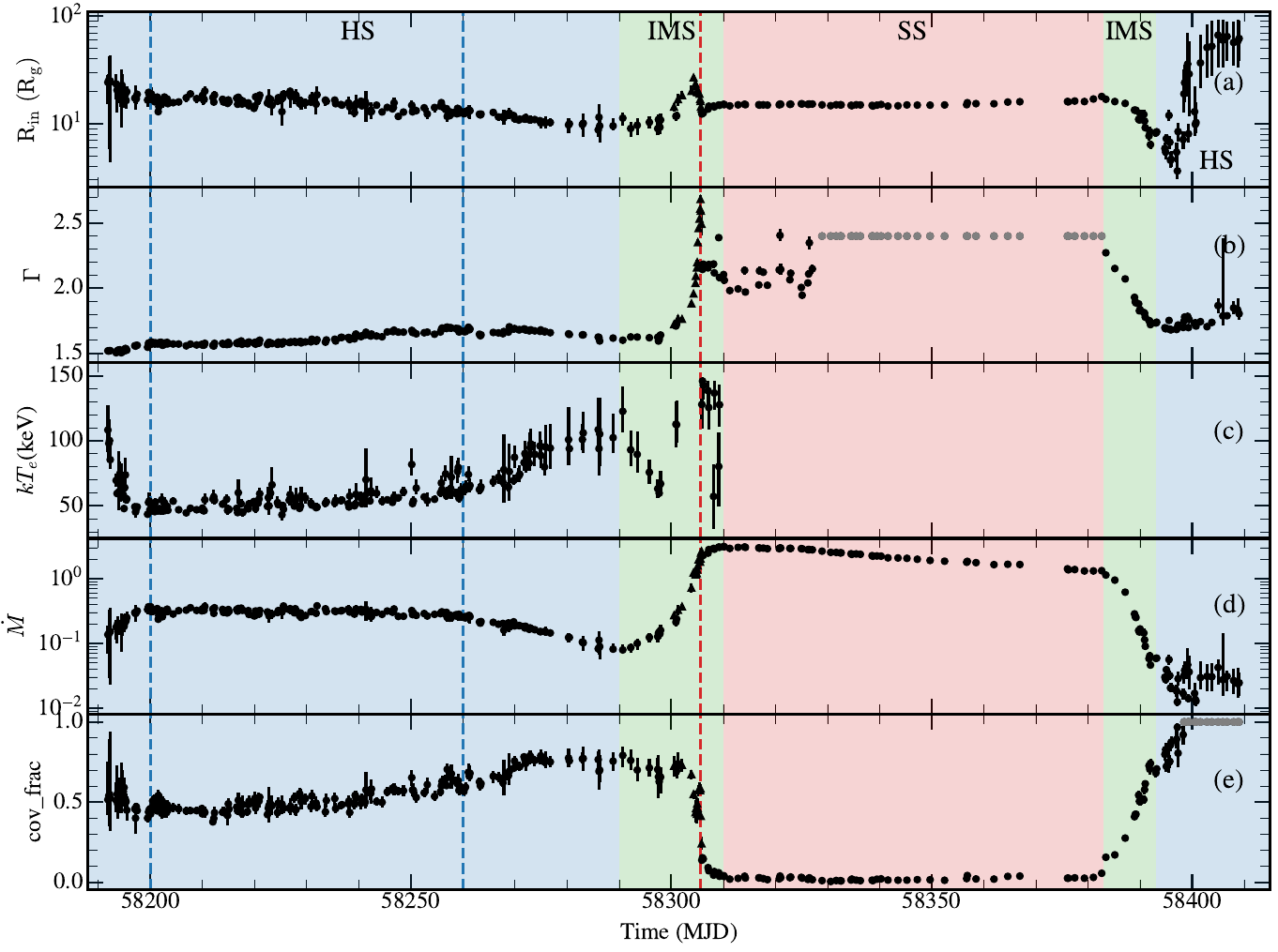}
\caption{Time evolution of spectral parameters. From top to bottom: the inner disk radius $R_\mathrm{{in}}$, the photon index $\Gamma$, the electron temperature $kT_e$, the mass accretion rate $\dot{M}$, the cover fraction \texttt{cov\_frac}. The dots are derived from $Insight$-HXMT data and the triangles are derived from $NICER$ data. The grey dots represent the fixed values. The colors of the background and dashed lines are the same as in \autoref{fig:hidall}.} 
\label{fig:sedpara}
\end{figure*}

\begin{figure*}
\centering
\includegraphics[width=1\textwidth]{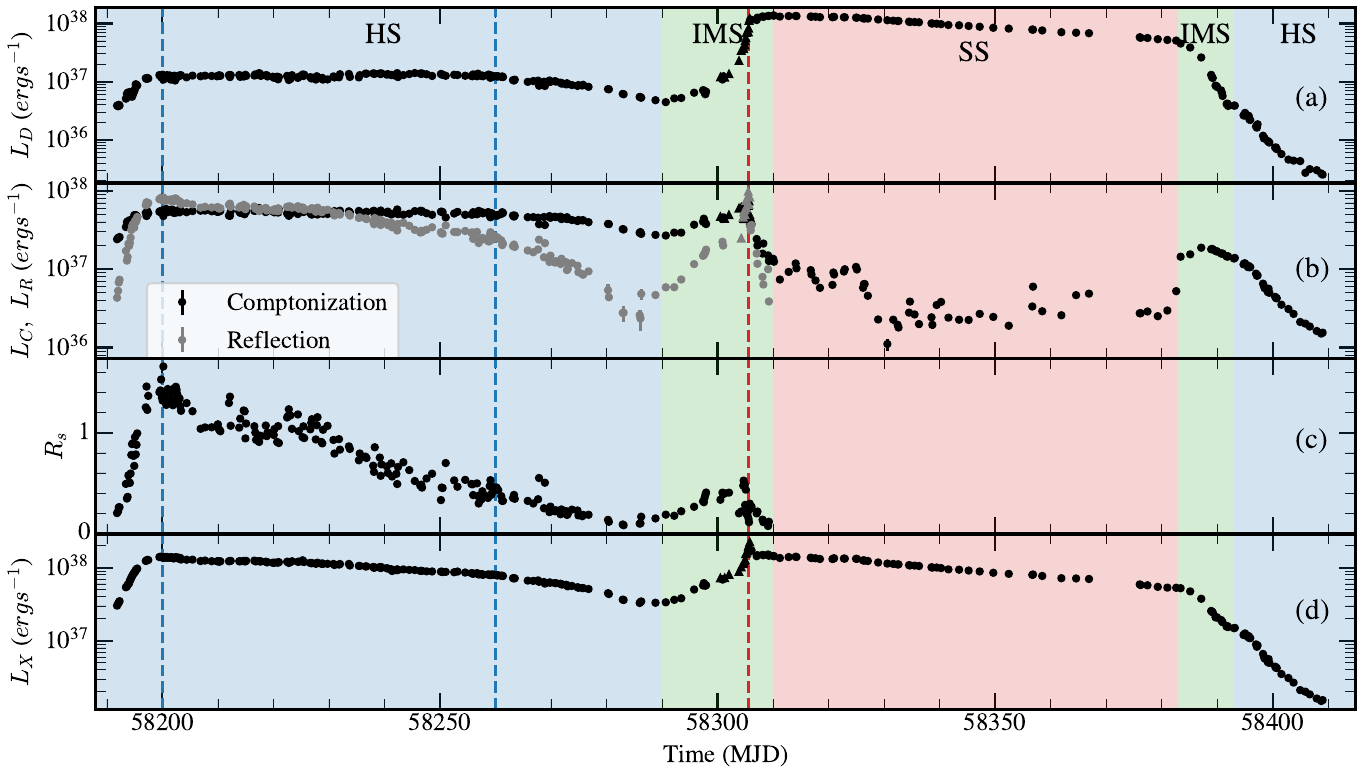}
\caption{Time evolution of X-ray luminosities in 0.1--100 keV of different spectral components: the disk luminosity (a), the Comptonization luminosity and the reflection luminosity (b) and total luminosity (d). We present the reflection strength $R_{s}$ at (c), calculated as the ratio of $L\mathrm{_{R}}$ to $L\mathrm{_{C}}$ in 20--40 keV energy band. The dots are derived using $Insight$-HXMT data and the triangles are derived using NICER data. The colors of the background and dashed lines are the same as in \autoref{fig:hidall}. } 
\label{fig:lumin}
\end{figure*}

Based on spectral analysis, we present the evolution of the main parameters of the models in \autoref{fig:sedpara}.
The inner disk radius $R_\mathrm{{in}}$ steadily decreases during the HS. However, it increases during the IMS and then decreases to the stable value of SS.
The photon index $\Gamma$ manifests different growth rates in the rise and plateau phases of HS, followed by a decrease during the bright decline phase and a rapid increase during the IMS.
The electron temperature $kT_e$ rapidly decreases in the rise phase of HS, then gradually increases until the end of HS. During the IMS, it goes down and up. The mass accretion rate $\dot{M}$ increases in the rise phase of HS and remains almost stable in the plateau phase, then decreases in the bright decline phase. Subsequently, it increases during the IMS.
The \texttt{cov\_frac} which represents the ratio of Comptonized seed photons slowly rises in the HS and rapidly declines in the IMS.

During the SS, the $\dot{M}$ decreases exponentially accompanied by a relatively constant $R_\mathrm{{in}}$. $\Gamma$ at an early stage of SS is lower than that in SIMS, which demonstrates that there is a strong hard X-ray tail in the $Insight$-HXMT spectra \citep[see also ][]{Fabian2020,Mummery2024}. The $\Gamma$ is fixed at $2.4$ for the rest of SS. During the decaying IMS, $\Gamma$ gradually drops to the HS level $\sim$ 1.7. The \texttt{cov\_frac} remains at almost zero value during the SS and increases to the HS level in the decaying IMS.
The evolution of the spectral parameters in different spectral states is also reported in \citet{YouB2021,Peng2023,FanNY2024}. 

The X-ray luminosity of different spectral components calculated in the energy range of 0.1--100 keV is presented in \autoref{fig:lumin}, including the total X-ray luminosity $L\mathrm{_{X}}$, the reflection luminosity $L\mathrm{_{R}}$, the Compton luminosity $L\mathrm{_{C}}$, and the disk luminosity $L\mathrm{_{D}}$. In the rise phase of HS, the luminosites of all spectral components increase. Subsequently, $L\mathrm{_{R}}$ and $L\mathrm{_{X}}$ decrease until the end of HS. 
We calculated the ratio of $L\mathrm{_{R}}$ to $L\mathrm{_{C}}$ in 20--40 keV energy band as the reflection strength $R_{s}$ \citep{Dauser2016}, which increases to the peak synchronously with $L\mathrm{_{R}}$ during the rise phase of HS and then decreases during the plateau and bright decline phases.
Notably, $L\mathrm{_{C}}$ and $L\mathrm{_{D}}$ remain stable during the plateau phase and then decrease in the bright decline phase of HS. During the rising IMS, there is a significant increase in luminosities of different spectral components, particularly before the period of transient jet ejection (\autoref{fig:lumin}). $L\mathrm{_{D}}$ exhibits an exponential decline during the SS, then continues to decrease at a steeper rate during the decaying IMS and HS, while $L\mathrm{_{C}}$ experiences a small flare (\autoref{fig:lumin}).

\subsubsection{Relationships between $L_\mathrm{X}$ and spectral parameters}

\begin{figure*}
\centering
\includegraphics[width=1\textwidth]{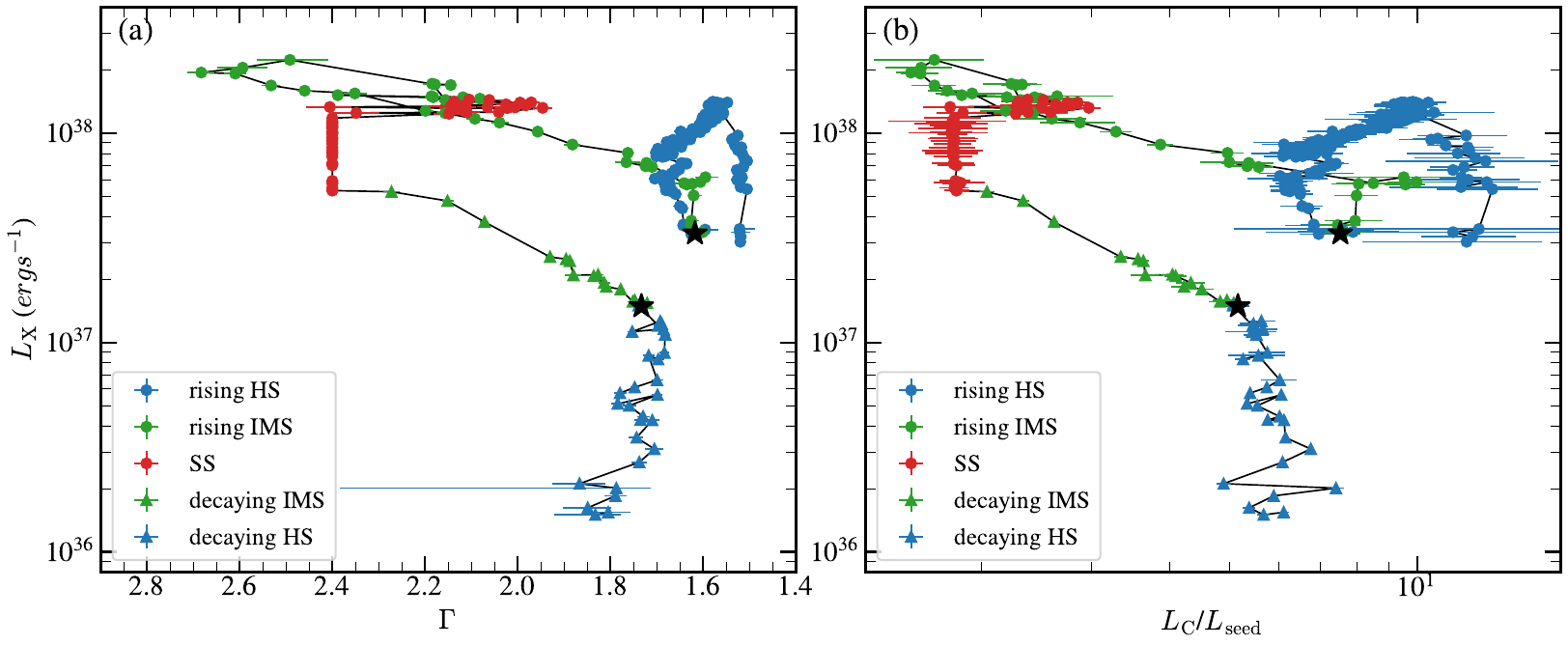}
\caption{(a) The relationship between photon index $\Gamma$ and total luminosity $L_\mathrm{X}$ during the 2018 outburst. Blue dots, green dots, red dots, green triangles and blue triangles respectively indicate rising hard state and intermediate state, soft state, decaying hard state and intermediate state. The black stars mark the transition luminosities between HS and IMS in the outburst rise and decay phases. Arrows indicate the direction of evolution. 
(b) The relationship between the ratio of Comptonization luminosity $L\mathrm{_{C}}$ to seed photons luminosity $L\mathrm{_{seed}}$ and total luminosity $L_\mathrm{X}$ during the 2018 outburst. Symbols and colors are the same as in panel (a).}
\label{fig:LgammaLcLd}
\end{figure*}

\begin{figure*}
\centering
\includegraphics[width=1\textwidth]{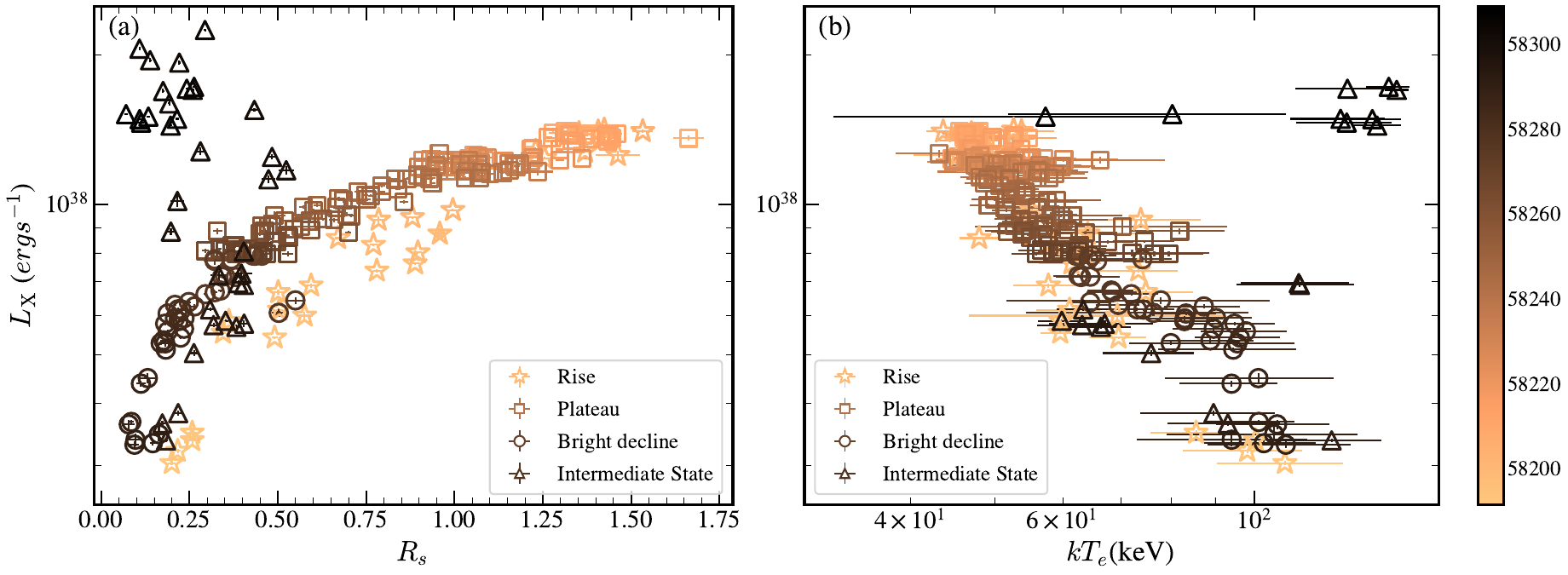}
\caption{(a) The relationship between the reflection strength $R_s$ and total luminosity $L_\mathrm{X}$ during the hard/hard-intermediate state. The stars, squares, dots and triangles respectively indicate the rise, plateau and bright decline phases in the hard state and intermediate state. 
(b) The relationship between the electron temperature $kT_e$ and total luminosity $L_\mathrm{X}$ during the hard/hard-intermediate state. Symbols and colors are the same as in panel (a). The color bar on the right corresponds to the observation time. }
\label{fig:LkTeRf}
\end{figure*}

We then investigated the relationships between X-ray luminosity and the spectral parameters. The relationship between photon index $\Gamma$ and total X-ray luminosity $L_\mathrm{X}$ is presented in \autoref{fig:LgammaLcLd}a. It is evident that $L_\mathrm{X}$ and $\Gamma$ show almost identical ``$\mathrm{q}$''-like evolution with HID even the small ``$\eta$''-like track during the rising HS. The reflection component during the HS of MAXI J1820+070 is very strong, especially during the plateau phase, which is almost the strongest among BHXRBs at the same luminosity \citep{YouB2023ApJ}. Notably, the slope of the reflection spectra in the 2--10 keV energy range are always steeper than that of the Compton spectra (\autoref{fig:modelref}), resulting in a larger $\Gamma$ and also hardness. So the photon index (also hardness ratio) during the plateau phase is seriously influenced by the reflection component. Especially when the reflection component surpasses the Compton component below the energy of the reflection hump in the spectra, the photon index is determined by the reflection component rather than the Compton component (\autoref{fig:modelref}). So the distinctive evolution of reflection component leads to the unique ``$\eta$''-like track in $L_\mathrm{X}$--$\Gamma$ and HID during the rising HS.

The relationships between total X-ray luminosity $L_\mathrm{X}$ and both the reflection strength $R_s$ and electron temperature $kT_e$ are respectively shown in \autoref{fig:LkTeRf}a and \autoref{fig:LkTeRf}b. During the rising HS and HIMS, a positive correlation is observed between $L_\mathrm{X}$ and $R_s$, which is not hold in the SIMS \citep[see also ][]{YouB2023ApJ}. Concurrently, $kT_e$ demonstrates a negative correlation with $L_\mathrm{X}$ during both the HS and HIMS \citep[e.g. ][]{YanZ2020,YouB2023ApJ}. The observations in SIMS deviates from this relationship.

\subsection{Type-C QPO evolution}
\begin{figure*}
\centering
\includegraphics[width=1\textwidth]{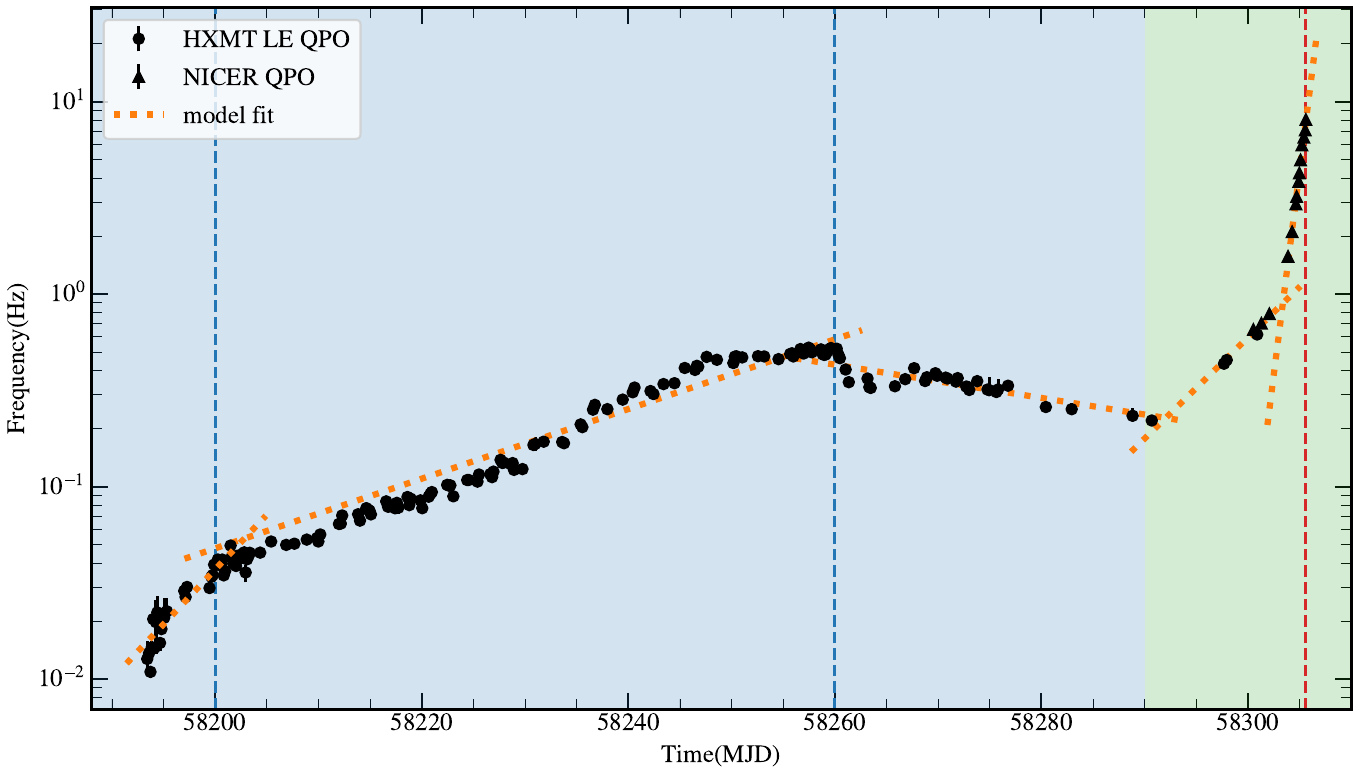}
\caption{Time evolution of the characteristic frequencies of the type-C QPO. The dots are obtained from $Insight$-HXMT LE data and triangles are from $NICER$ data. The orange dotted lines represent the best-fitting exponential functions applied to the data of different phases: the rise, plateau, and bright decline phases in hard state, and two segments separated on MJD 58303 during hard intermediate state. 
The colors of the background and dashed lines are the same as in \autoref{fig:hidall}.}
\label{fig:qpobln}
\end{figure*}

\begin{figure*}
\centering
\includegraphics[width=1\textwidth]{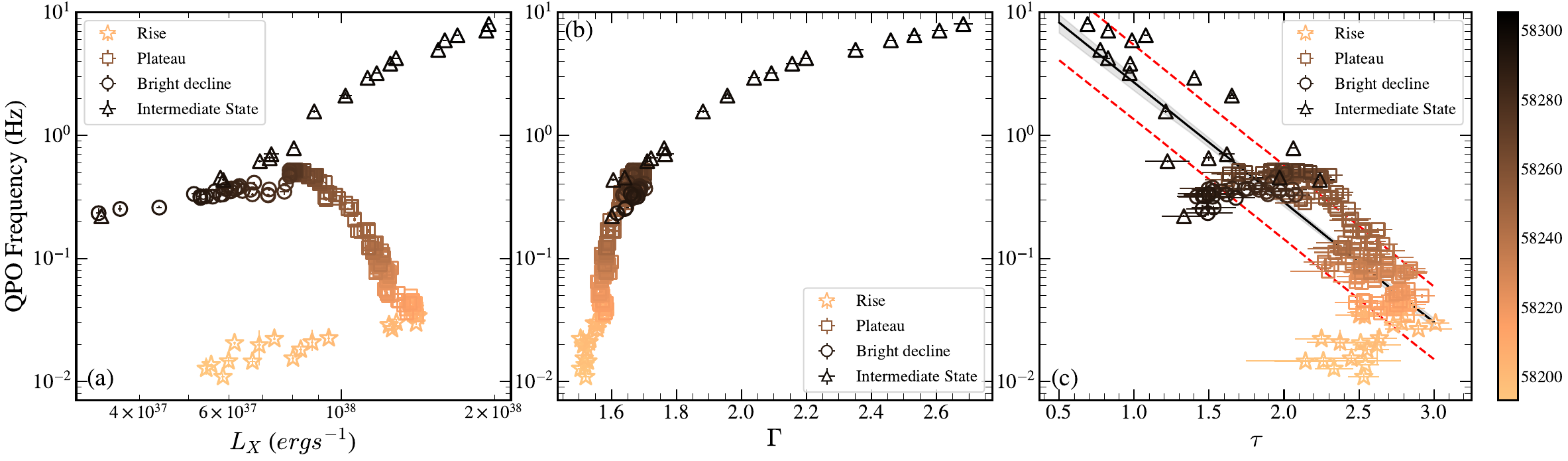}
\caption{(a) The relationship between type-C QPO frequency and total luminosity $L_\mathrm{X}$. (b) The relationship between type-C QPO frequency and photon index $\Gamma$. (c) The relationship between type-C QPO frequency and optical depth $\mathrm{\tau}$. The black line indicates the best-fitting linear function. The gray shaded area represents the error region for the slope and offset. The red dashed lines indicate the scattering range of the data. Symbols and colors are the same as in \autoref{fig:LkTeRf}. The color bar on the right corresponds to the observation time. }
\label{fig:qpoLGTao}
\end{figure*}

The evolution of centroid frequency of type-C QPO is present in \autoref{fig:qpobln}. We could detect the type-C QPO from around MJD 58193.42 until MJD 58305.61 when the source reaches the SIMS. The centroid frequency of QPO $\nu_\mathrm{C,QPO}$ ranges from 0.01 to 8.05Hz and the values obtained from the same exposure ID in different detectors (LE, ME and HE) are nearly identical. The $\nu_\mathrm{C,QPO}$ evolution obtained from $Insight$-HXMT is almost the same as that from $NICER$ data \citep{Stiele2020}.

The QPO frequency apparently changes at different rates during different phases (\autoref{fig:qpobln}). \citet{Buisson2018} has claimed that the QPO frequency exponentially increases with time. We also used exponential function $\nu_\mathrm{C,QPO}=p_{0} e^{(t-t_{0} )/\tau }$ to fit the $\nu_\mathrm{C,QPO}$ at different phases. The best-fitting $e$-folding time scales $\tau$ are 7.63 $\pm$ 0.78, 23.99 $\pm$ 0.65, -50.27 $\pm$ 7.86 days for the rise, plateau and bright decline phases of HS, respectively (\autoref{fig:qpobln}). The $\nu_\mathrm{C,QPO}$ evolution during the IMS obviously experiences two different increasing rates with $e$-folding time scales of 8.28 $\pm$ 0.76, and 1.02 $\pm$ 0.05 days \autoref{fig:qpobln}.

We also performed timing analysis for the $Insight$-HXMT data during the decaying IMS and HS. However no significant QPO signals were detected. We noticed that there are some QPO detections with $NICER$ data reported in \citet{Stiele2020}. However, only six QPO signals are considered to be significant ($>3\sigma$). Their frequency and fractional RMS vary chaotically. Consequently, it is difficult to conclusively determine the type and evolution of the detected QPOs during the decaying IMS and HS.

We then investigated the relationships between type-C QPO frequency $\nu_\mathrm{C, QPO}$ and spectral parameters during the rising HS and HIMS (\autoref{fig:qpoLGTao}). The $\nu_\mathrm{C, QPO}$ exhibits a complex correlation with the total X-ray luminosity $L_\mathrm{X}$. Two positive correlations parallelly exist in the rise and bright decline phases of HS and HIMS. However, a negative correlation is observed in the plateau phase of HS (\autoref{fig:qpoLGTao}a). 
The $\nu_\mathrm{C, QPO}$ and photon index $\Gamma$ exhibit an overall positive correlation in the HS and HIMS, with varying slopes across different phases (\autoref{fig:qpoLGTao}b). It appears that the slopes during the rise, plateau, and bright decline phases of HS and HIMS differ. We also calculated the optical depth $\mathrm{\tau}$ of Compton component from $kT_e$ and $\Gamma$ according to the equation in \citep{Zdziarski1996}. As present in \autoref{fig:qpoLGTao}c, the $\nu_\mathrm{C, QPO}$ and $\mathrm{\tau}$ roughly follow a negative correlation except the observations in the rise phase of HS. The best-fitting slope of this correlation is obtained as $-0.96\pm 0.05$ using the Bayesian method in \citet{Buchner2021}. However, the correlation displays a large scatter, with a value of $0.30\pm 0.02$.

\subsection{Dramatic spectral and timing changes around the transient jet ejection}
\label{sec:4.4}

 \begin{figure}
\centering
\includegraphics[width=0.5\textwidth]{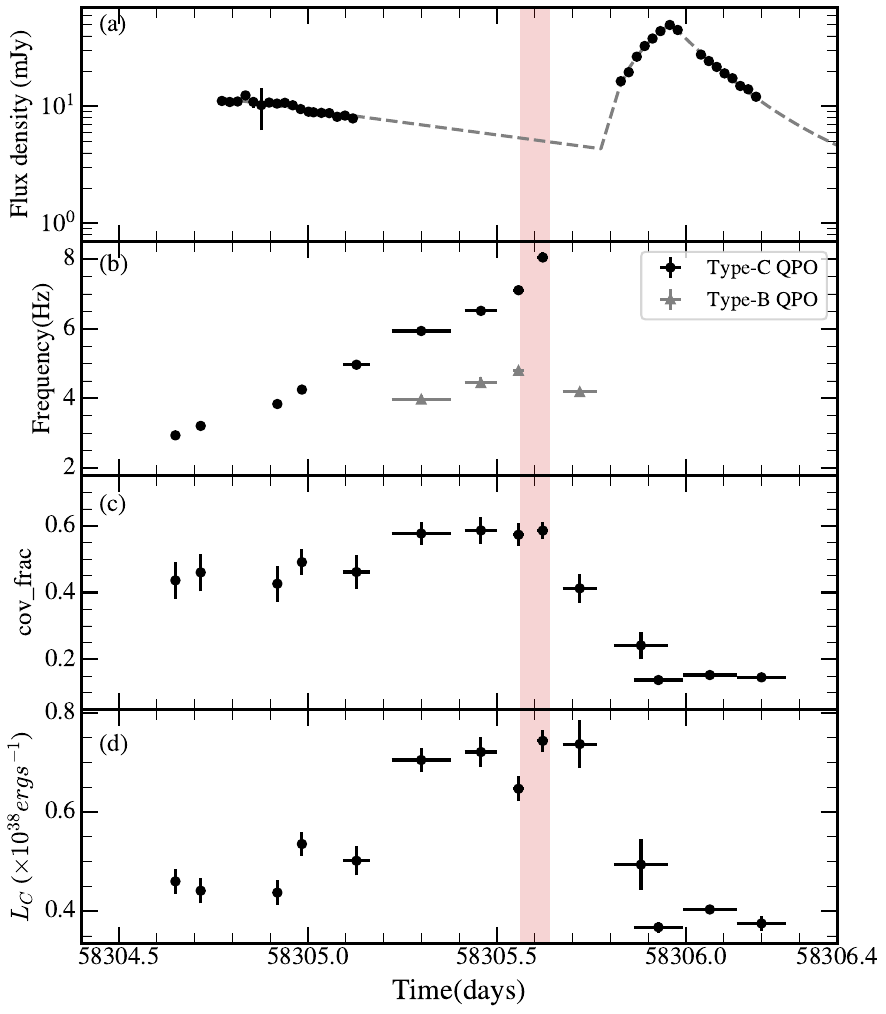}
\caption{(a) AMI-LA radio light curve (15.5 Ghz) from \citet{Bright2020} around the transient jet ejection. The grey dashed line indicates the best-fitting model from \citet{Homan2020}: an exponential function for the baseline decay, an additional linear function for the radio flare rise, and an additional power-law function for the radio flare decay phase.
(b) The concurrent evolution of the centroid frequency of type-C and type-B QPOs. The red shadows mark the transient jet ejection time on MJD 58305.60 $\pm$ 0.04 according to \citep{Wood2021}.
(c) The concurrent evolution of the cover fraction \texttt{cov\_frac}. 
(d) The concurrent evolution of the Comptonization luminosity $L\mathrm{_{C}}$ (0.1--100 keV). }
\label{fig:eject}
\end{figure}

Dramatic changes of X-ray spectral and timing coincides with transient jet ejections around MJD 58306 \citep{Bright2020, Homan2020, Wood2021}.
Due to the extensive coverage of $NICER$ observation, we are able to investigate the spectral and timing variation before and after the ejection. 
We divided the two observations (ObsID 1200120196 and 1200120197) around transient jet ejection into 12 segments. The parameters of QPOs detected in each segment are listed in \autoref{Tab:pds196197}. The frequency of type-C QPO increases until the last detected frequency reaches 8.05 Hz. Notably, the last detection of type-C QPO just coincides with the transient jet ejection on MJD 58305.60 $\pm$ 0.04 (\autoref{fig:eject}).
Additionally, a type-B QPO around 4 Hz emerges roughly 0.4 days before this ejection. It is first detected on MJD 58305.22 and persists for approximately 13 hours. Although the type-B QPOs are not prominently visible in the dynamic PDSs prior to the ejection \citep[see ][]{Homan2020}, they are significantly detected in the averaged PDS of certain segments  (\autoref{Tab:pds196197}).

We also generated the energy spectra for each segment. In spectral fitting, we fixed the electron number density $\log N$, iron abundance $A\mathrm{_{Fe}}$ and ionization parameter $\mathrm{log}~\xi$. After the transient jet ejection, $L\mathrm{_{C}}$ drops by a factor of 2 within 8 hours, and \texttt{cov\_frac} rapidly decreases to the level of SS.

\section{Discussion}
\subsection{The ``q''-like HID and the evolution of $L_\mathrm{X}$-$\Gamma$}

Similar evolution tracks are anticipated in both the $L_\mathrm{X}$--$\Gamma$ and HID, as the calculated hardness (3--10keV/1--3keV) depends on the slope of the X-ray spectrum, also known as the photon index. \autoref{fig:LgammaLcLd}a displays a ``q"-like track between $L_\mathrm{X}$ and $\Gamma$ when all data from the entire outburst are included, and a small ``$\eta$"-like track during the rising HS, which is similar to HID (\autoref{fig:hidall}). 
 We have also gathered data from several outbursts with extensive observations that span the full duration of the outburst from the literature \citep{Sobczak2000,Debnath2008, McClintock2009}. We found that $L_\mathrm{X}$ and $\Gamma$ follow a roughly ``q"-like track in all three outbursts similar to the HID (see \autoref{fig:others}). The data in the hard state (right part of the ``q") roughly follows a ``v"-like correlation, aligning with previous work \citep[e.g. ][]{YangQX2015,YanZ2020}. The significant scatter of the right branch of ``v"-like correlation \citep[e.g. Fig.2 in][]{YanZ2020} is evidently due to the different bend points at the upper right of the ``q'' track in different outbursts, i.e., the luminosity where the HS can reach.

In the case of thermal Comptonization scenario, the photon index $\Gamma$ monotonically decreases with $L_\mathrm{corona}/L_\mathrm{seed}$ \citep[e.g.][]{Haardt1991,Beloborodov1999}. 
The relationship in this work indeed shows the negative correlation anticipated by the model from \citet{Beloborodov1999} (see \autoref{fig:LcLseedgamma}). We then present the relationship between the $L_\mathrm{X}$ and the ratio of $L\mathrm{_{C}}$ to $L\mathrm{_{seed}}$  in \autoref{fig:LgammaLcLd}b. The $L\mathrm{_{seed}}$ is calculated by multiplying the intrinsic $L\mathrm{_{D}}$ with \texttt{cov\_frac}. This diagram also presents a ``q''-like pattern similar to HID and $L_{X}$-$\Gamma$ including the ``$\eta$''-like track during the rising HS, which is consistent with the expectation of thermal Comptonization scenario \citep{Beloborodov1999}.

However, the reflection radiation is strong during the rising HS. So the seed photons should be dominated by the reprocessed emission from the accretion disk rather than the intrinsic disk emission \citep{Zdziarski1999,Beloborodov1999}. We cannot distinguish the reprocessed and intrinsic disk emission from current spectral fitting. In the convolution model \texttt{thcomp*kerrd} we used in the rising HS, the disk emission serves as seed photons for Compton scattering, therefore, the best-fitting parameters of disk component may not represent the true standard thin disk.

The H-S transition luminosity during the outburst rise is usually several times larger than the S-H transition luminosity during the outburst decay\citep[e.g.][]{YuWF2009,Tetarenko2016}, even the hardness ratio or photon index is similar. This phenomenon is called hysteresis effect, which is universal in outbursts experiencing state transitions \citep[e.g.][]{Maccarone2003}. The transition luminosities of  H-S and S-H correspond to the last observation in rising HS and the first observation in decaying HS. For MAXI J1820+070, the H-S transition luminosity is 3.32 $\times10^{37}$ erg s$^{-1}$, which is almost two times higher than the S-H transition luminosity (1.49 $\times10^{37}$ erg s$^{-1}$, see also \autoref{fig:LgammaLcLd}). The Compton luminosity of H-S transition is also roughly two times higher than that of S-H transition (2.73$\times10^{37}$ and 1.37$\times10^{37}$ erg s$^{-1}$), while the disk luminosities are similar (1.17$\times10^{36}$ and 1.23$\times10^{36}$). 
So the discrepancy of the transition luminosity is dominated by the Compton component. 

Hot accretion flow is the predominant model to produce the Compton component during the HS \cite[see reviews by][]{Yuanf2014,LiuBF2022}. In such a framework, the hot accretion flows can only exist below the critical luminosity $L<\alpha^{2}L_\mathrm{Edd}$ \citep[e.g.][]{Esin1997,XieFG2012}, where the $\alpha$ is the viscosity parameter of hot accretion flow and the $L_\mathrm{Edd}$ is the Eddington luminosity. So an increasing $\alpha$ can maintain the hot accretion flow at a high luminosity regime, to make the H-S transition luminosity higher \citep{Begelman2014,CaoXW2016,LiJQ2023}. The enhancement of the magnetic field and/or outflow is able to increase $\alpha$ \citep{BaiXJ2013,CaoXW2016}. 
Non-stationary accretion flow has been proposed to drive the H-S transition luminosity during outburst rise, i.e. the transition luminosity is higher when the mass accretion rate increases faster \citep{YuWF2004,YuWF2009}. As a result, the increasing $\alpha$ may play a similar role as the proposed non-stationary accretion flow in \citet{YuWF2009}. 

Jet or its base is also an alternative origin of the Compton emission. Some X-ray spectral and timing properties during the state transition can be well explained by jet-like corona (see the discussions in subsequent sections). \citet{Marcel2019} has fine-tuned two independent parameters $r_\mathrm{J}$ and $\dot{m}_{in}$ of the jet-accretion disk model to replicate the ``q''-like evolutionary track, where $r_\mathrm{J}$ is the transition radius between jet and accretion disk, $\dot{m}_{in}$ is the mass accretion at the inner disk radius. This model is quite similar to the truncated disk model \citep{Esin1997,Done2007}. However, it is still unknown what mechanism drives the evolution of $r_\mathrm{J}$.  

\subsection{Inner radius of accretion disk/reflected component} 
\label{5.2}

We used the \texttt{kerrd} to fit the thermal component in HS, IMS and SS and acquired the evolution of $R_{\mathrm{in}}$ (see \autoref{fig:sedpara}a). In an optically-thick disk, the effective temperature of the blackbody emission as a given radius $r$ will be $T_{\mathrm{eff}} (r)=\left [ \frac{3GM\dot{M}  }{8\pi \sigma r^{3} }  \left ( 1-\sqrt{\frac{r_{\mathrm{in}}}{r} }  \right ) \right ] ^{\frac{1}{4} } $ \citep{1973A&A....24..337S, Makishima1986}, where $G$ is the constant of gravity, $M$ is the BH mass, $\dot{M}$ is the mass accretion rate and $r_{\mathrm{in}}$ is the inner disk radius considering the effect of inner boundary condition. Then, the maximum disk temperature is given as $T_{\mathrm{max}} = (3GM\dot{M}/8\pi \sigma r_{\mathrm{in}}^{3} )^{1/4}\times 6^{3/2} \times 7^{-7/4} $\citep{Kubota1998}. Therefore, the disk luminosity is $L_{\mathrm{D}} \sim 4\pi R_{\mathrm{in}}^{2} \sigma T_{\mathrm{max}}^{4} = (3GM\dot{M} R_{\mathrm{in}}^{2}/2r_{\mathrm{in}}^{3})\times 6^{6}\times 7^{-7}$, where $R_{\mathrm{in}}$ is our spectral fitting result. So we can obtain the true inner disk radius from spectra fitting result, since $r_{\mathrm{in}}  = 0.44\left (\dot{M}c^{2}R_{\mathrm{in}}^{2} /{L_{\mathrm{D}}}\right )$ ( in units of $R_\mathrm{g}$). During the SS, the nearly constant $R_{\mathrm{in}}$ and similar evolution of $L_{\mathrm{D}}$ and $\dot{M}$ result to an almost unchanged $r_{\mathrm{in}} \sim 7 R_{\mathrm{g}}$, which is a little bit larger than the $R_{\mathrm{ISCO}}$ of a low-spinning BH \citep{2021ApJ...916..108Z,GuanJ2021}. Therefore, the exponential decreasing of $\dot{M}$ and constant $r_{\mathrm{in}}$ during the SS (\autoref{fig:sedpara}) is consistent with the evolution of a standard accretion disk \citep{1973A&A....24..337S}.

In the spectral fitting of the rising HS and IMS, we employed the relativistic reflection model \texttt{relxillCp} and linked the inner radius of the reflected component to the $R_\mathrm{in}$. We then calculated the corresponding $r_\mathrm{in}$. During the HS, the $r_\mathrm{in}$ continuously decreases, and then increases after entering the IMS (\autoref{fig:compareRin}). The lowest $r_\mathrm{in}$ around MJD 58290 is $4.5 R_\mathrm{g}$. It is odd that the inner disk radius increases following the transition to IMS, yet the values observed afterward are smaller compared to those during the SS (\autoref{fig:sedpara}). We also used another model to fit the broadband X-ray spectra. Although the values of inner disk radius during the HS and IMS are different from \texttt{kerrd}, the evolutions are similar (see \autoref{app:appa}). 
Especially the inner radius experiences a dip during the IMS, the value of which is lower than that in the SS \citep[\autoref{fig:compareRin}, see also][]{Motta2009}.

The observational results present a significant challenge to our current understanding of the accretion flow evolution during the IMS, a topic that remains highly debated. On one hand, current spectral models may require refinement to accurately describe the broad-band X-ray spectra observed around IMS. On the other hand, the thermal component may not be an accretion disk. Instead, the $R_\mathrm{in}$ serves as a proxy for the emission area of thermal component rather than the inner radius of accretion disk, as $R_\mathrm{in}$ is also influenced by the total flux of the thermal component. However, the nature of this component remains puzzling, especially if its size is smaller than the $R_\mathrm{ISCO}$ of the standard accretion disk. One possible candidate is the presence of cold clumps embedded within a hot accretion flow. This two-phase accretion flow at bright hard state and/or IMS has been proposed in many previous studies \citep[e.g.][]{Krolik1998,WangJM2012,XieFG2012,Liska2022}. Nevertheless, the radiative properties of such an accretion flow, especially the characteristics of the cold clumps, remain poorly understood. Further investigation is required to clarify whether it can account for the X-ray spectral evolution around IMS. 

We also noticed that the evolution of $R_\mathrm{in}$ is roughly consistent with the time lag between the soft and hard X-ray variability \citep[e.g.][]{WangJY2021,DeMarco2021,Echiburu2024}. The soft lag is interpreted as the reverberation lag between the corona and the reflected component \citep[e.g.]{Kara2019,WangJY2021,WangJY2022}. The decreasing of soft lag during the HS agrees with the expected reflection from a truncated accretion disk or a contracting corona \citep{Kara2019,DeMarco2021}. The increase of soft lag during the IMS is proposed to be related to an expanding of jet or jet-like corona \citep{WangJY2021,DeMarco2021}. A jet-like corona is also proposed to explain the other spectral and timing properties in BHXRBs \citep[see also][]{Kylafis2008,YouB2021,CaoZ2022,Mendez2022,WangJY2022}. Under this scenario, a jet-like corona undergoes vertical contraction and expansion, resulting in the observed decrease and increase of the soft time lag and the irradiated area.

\subsection{Type-C QPO frequency, corona geometry and Comptonization}

The type-C QPO is a predominant characteristic detected in the HS and HIMS, the oscillated photons of which are generally believed to originate from the component producing the Compton emission \citep[e.g.][]{Lee1998,Psaltis2000,Bellavita2022,Ma2021,gao2023}. So it is important to investigate the QPO properties and the spectral parameters of Compton component.
The correlation between the QPO frequency $\nu_\mathrm{C,QPO}$ and the photon index $\Gamma$ of the Compton component has been widely studied \citep[e.g.][]{Vignarca2003,Shaposhnikov2009,Stiele2013,WangPJ2022}. The $\nu_\mathrm{C,QPO}$ usually positively correlates with the $\Gamma$ over two orders of magnitude frequency range, and the $\Gamma$ becomes saturated above a certain frequency in few sources \citep[e.g.][]{Vignarca2003,Shaposhnikov2009}. 
It is generally believed that the mass accretion rate drives the evolution of both X-ray spectral and timing properties of BHXRB\citep[see reviews in ][]{Done2007,Gilfanov2010}. The $\Gamma$ and the $\nu_\mathrm{C,QPO}$ both increase with X-ray luminosity, resulting in a positive correlation \citep[e.g.][]{Revnivtsev2001,Mereminskiy2019,WangPJ2022}. However, $\nu_\mathrm{C,QPO}$ in MAXI J1820+070 displays a complex correlation with $L_\mathrm{X}$ (\autoref{fig:qpoLGTao}a), and $\Gamma$ exhibits a ``$\eta$''-like track along $L_\mathrm{X}$ during HS (\autoref{fig:LgammaLcLd}). These results suggest that the mass accretion rate is not the only factor influencing the evolution of the spectral and timing properties of the corona.

MAXI J1820+070 also shows a positive correlation between $\nu_\mathrm{C,QPO}$ and $\Gamma$ with varying slopes across different phases of HS and HIMS (\autoref{fig:qpoLGTao}b). The $\nu_\mathrm{C,QPO}$ and $\tau$ roughly exhibit a negative correlation with a large scatter (\autoref{fig:qpoLGTao}c). These correlations provide compelling evidence that the Comptonization is the driving force behind the evolution of the QPO frequency. Recently, some models have been proposed to explain the properties of the type-C QPO based on the Comptonization process \citep[e.g.][]{Bellavita2022,Mastichiadis2022}. Specifically, the interplay between thermal and Compton components, which influences optical depth and electron temperature during Comptonization, can produce an oscillation roughly consistent with type-C QPO \citep{Mastichiadis2022}.  

In some other models, the QPO frequency is usually thought to inversely correlate to the characteristic scale of the corona and/or the inner radius of the accretion disk \citep[e.g.][]{Ingram2009,Cabanac2010,Marcel2020,Rawat2023,Motta2024}. 
So the increasing of QPO frequency during the rise and plateau phases of HS (\autoref{fig:qpobln}) can be accounted for the shrinking of the inner radius of a truncated accretion disk \citep{DeMarco2021,Zdziarski2021} or a contracting corona \citep{Kara2019,WangJY2021,WangJY2022}. However,
the observed decrease in QPO frequency during the bright decline, and its subsequent increase during the IMS (\autoref{fig:qpobln}), is challenging to be explained by the behavior of a contracting and then expanding jet-like corona inferred from the soft lag and X-ray spectral evolution \citep{WangJY2022}, and also contradicts with the evolution of inner disk radius under a truncated disk scenario. 

The QPO frequency may alternatively be proportional to a characteristic timescale for ejecting mass as a damped oscillator in a jet-like corona model. As a result, the $\nu_\mathrm{C,QPO} \propto R_{0}^{2}n_{e,0}$ \citep{Kylafis2008}, where the $R_{0}$ is the jet base, and $n_{e,0}$ is the electron number density of the jet base. The expanding jet-like corona during the IMS naturally causes the rapid increase of the QPO frequency. The optical depth in such corona should also be a function of $R_{0}$ and $n_{e,0}$. However, the specific relation between $\tau$ and $\Gamma$ in the jet-like corona should be investigated in order to quantitatively understand the correlations between $\nu_\mathrm{C,QPO}$ and $\tau$/$\Gamma$.

\subsection{The association between rapid state transition and transient jet ejection}
It is widely recognized that transient jet ejections are associated with X-ray spectral state transitions \citep{Fender2004}, wherein Compton-dominated spectra shift to thermal dominated spectra. The well-sampled observations of MAXI J1820+070 offer a unique opportunity to investigate the connection between ejection event and spectral state transition. Remarkably, the last detection of type-C QPO coincided precisely with the ejection time within one hour. Subsequent to this ejection, the Compton luminosity sharply decreased by a factor of two within 8 hours. So the observations of MAXI J1820+070 establish a coincidence between the corona and ejection within timescale of hours \citep[see also ][]{Homan2020}, which the corona is responsible for the Compton emission and the type-C QPO. One plausible scenario is that the corona or part of the corona is ejected, subsequently allowing the disk component to dominate the X-ray spectrum. The dominance of ions in the ejection composition further supports this hypothesis \citep{Zdziarski2024}. On the other hand, \citet{MaRC2023} also proposed a double corona model, while the larger corona transforms into a vertical jet-like corona at the ejection time.
Additionally, the corona transforming into an ejected jet has been proposed in the microquasar GRS 1915+105 \citep{Mirabel1998,Mendez2022}.
A similar scenario has also been proposed for the active galactic nuclei (AGNs) harboring super-massive BH, such as 3C 120 \citep{Marscher2002} and 3C 111 \citep{Chatterjee2011}, in which the observed X-ray dips are associated with the radio jet ejections.

\section*{Acknowledgements}
We would like to thanks the helpful discussion with Bei You, Fuguo Xie and Phil Uttley. This work made use of the data from Insight-HXMT mission, a project funded by China National Space Administration (CNSA) and the Chinese Academy of Sciences (CAS).
This work was supported in part by the Natural Science Foundation of China (grants 12373049, U1938114, 12361131579, U1838203 and 12373050).

\section*{Data Availability}
The \textit{Insight}-HXMT and $NICER$ data underlying this article are available in the public archive \url{http://archive.hxmt.cn/proposal} and \url{https://heasarc.gsfc.nasa.gov/docs/archive.html}.



\bibliographystyle{mnras}
\bibliography{J1820} 



\clearpage
\appendix

\section{Spectral fitting results of a replaced model for the rising HS}
\label{app:appa}

\begin{figure*}
\centering
\includegraphics[width=1\textwidth]{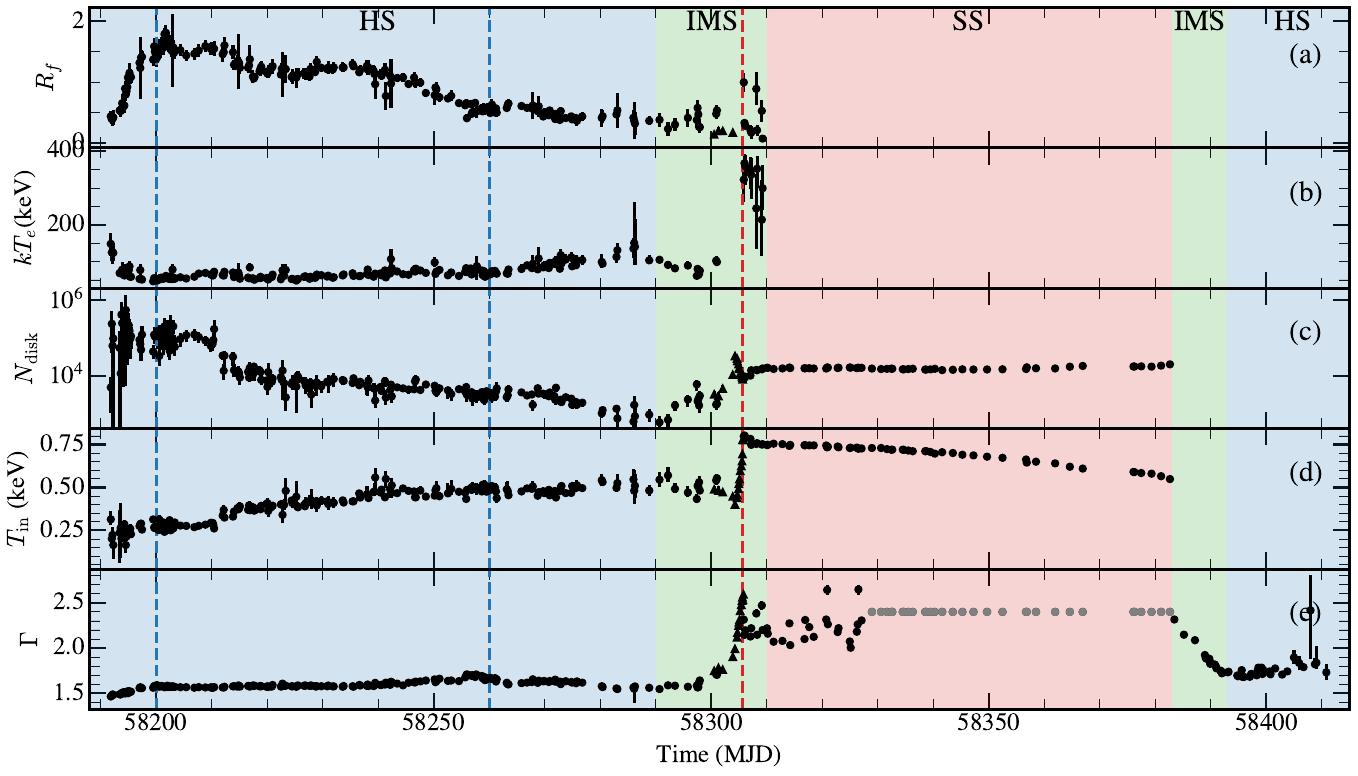}
\caption{Time evolution of spectral parameters of \texttt{constant*TBabs*(diskbb+relxillCp+xillverCp)}. From top to bottom: the reflection fraction $R_f$, the electron temperature $kT_e$, the normalization of the \texttt{diskbb} $N\mathrm{_{disk}}$, the inner disk temperature $T\mathrm{_{in}}$, the photon index $\Gamma$. The dots are derived from $Insight$-HXMT data and the triangles are derived from $NICER$ data. The grey dots represent the fixed values. The colors of the background and dashed lines are the same as in \autoref{fig:hidall}.} 
\label{fig:sedparaA}
\end{figure*}

\begin{figure*}
\centering
\includegraphics[width=1\textwidth]{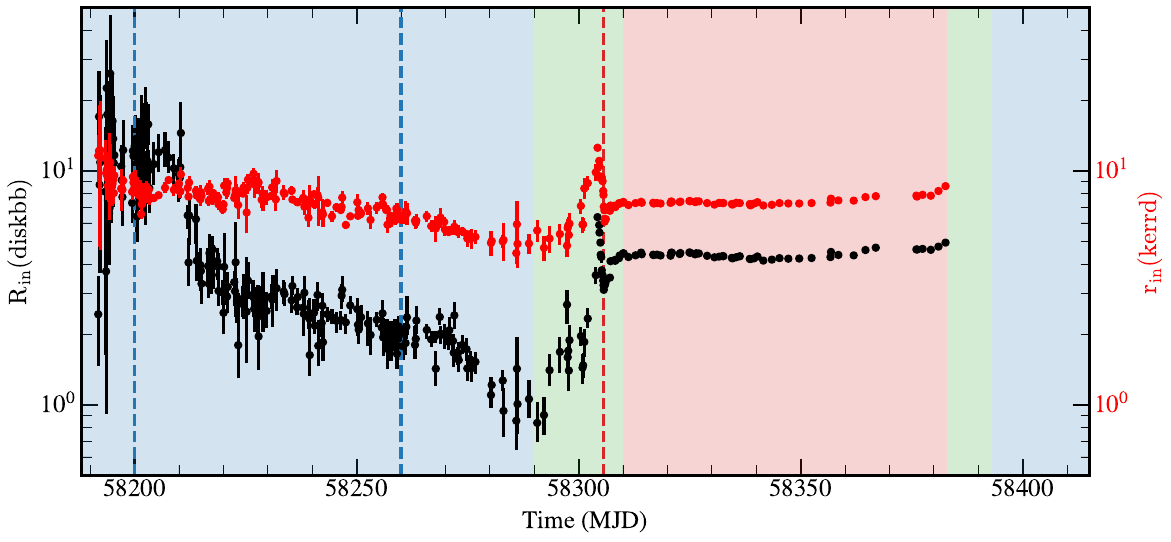}
\caption{We compared the inner disk radius estimated using the \texttt{diskbb} (\autoref{app:appa}) and \texttt{kerrd} (\autoref{3.4} and \autoref{5.2}) models. For the black dots, the model we use is \texttt{constant*TBabs*(diskbb+relxillCp+xillverCp)} during the HS and IMS and \texttt{constant*TBab*(diskbb+nthComp)} during the SS. For the red dots, the model is \texttt{constant*TBabs*(thcomp*kerrd+relxillCp+xillverCp)} during the HS and same model without \texttt{xillverCp} during the IMS and is \texttt{constant*TBabs*thcomp*kerrd} during SS. We have adopted a distance of 2.96kpc \citep{2020MNRAS.493L..81A}, an inclination angle of $63^{\circ } $ \citep{2020MNRAS.493L..81A}, and a black hole mass of 8.48$M_{\odot }$ \citep{2020ApJ...893L..37T}. 
Despite being derived from different models, the evolution of $\mathrm{r_{in}}$ appears roughly consistent.}
\label{fig:compareRin}
\end{figure*}

In order to verify the spectral parameters evolution, we have applied another model to describe the spectra observed in the rising HS and IMS, which is \texttt{constant*TBabs*(diskbb+relxillCp+xillverCp)}. The \texttt{diskbb} is a multi-blackbody component representing the accretion disk radiation. We used \texttt{relxillCp} to fit the Comptonization emission and the relativistic reflection emission. Furthermore, the same equivalent hydrogen column density, black hole spin, inclination angle and distance were maintained. However, the reflection fraction \texttt{refl\_frac} is free, which respresents the ratio of emitted photons from the corona illuminating the disk to the fraction of photons towards the observers \citep{Dauser2016}. It is worthy noting that we tied the normalization $N_\mathrm{disk}$ of the \texttt{diskbb} to the inner disk radius $R_\mathrm{in}$ (in units of $R_\mathrm{g}$) in \texttt{relxillCp} as 816$R_\mathrm{in}^2$ \citep{gao2023} since $N_\mathrm{disk} = (\frac{R_\mathrm{in}/1km}{D/10kpc} )^{2} \cos i$ \citep{Kubota1998,Basak2016}. 
During and after the soft state, we applied the model \texttt{constant*TBabs*(diskbb+nthComp)} to fit the broadband X-ray spectrum.

The evolution of the main parameters of these models is present in \autoref{fig:sedparaA}. We can find the same parameters have almost similar evolution to the results in the main text. The reflection fraction $R_f$ increases in the rise phase of HS and subsequently undergoes a gradual decline until the SS. The evolution of $R_{f}$ is almost identical to $R_{S}$ in \autoref{fig:lumin}, which demonstrates that they are tightly related \citep{Dauser2016}. The inner disk temperature $T\mathrm{_{in}}$ gradually increases until the decline phase of HS, and shows a rapid increase during the IMS, an exponential decrease during the SS.

The nearly constant $N_\mathrm{disk}$ and exponential decrease of $T_\mathrm{in}$ during the SS (\autoref{fig:sedparaA}) are also consistent with a standard accretion disk that the inner radius reaches the innermost stable circular orbit $R_\mathrm{ISCO}$ \citep{1973A&A....24..337S}. 
The $R_\mathrm{in}$ derived from \texttt{diskbb} is present in \autoref{fig:compareRin}. It is clear that the evolution of $R_\mathrm{in}$ derived from \texttt{diskbb} is consistent with that from $kerrd$. Especially, there is also a dip during the transition to the IMS, and the lowest $R_\mathrm{in}$ is lower than that in the SS.

\section{Comparison of spectral fittings between simultaneous NICER and $Insight$-HXMT observations}
\label{app:appb}

\begin{figure*}
\centering
\includegraphics[width=1\textwidth]{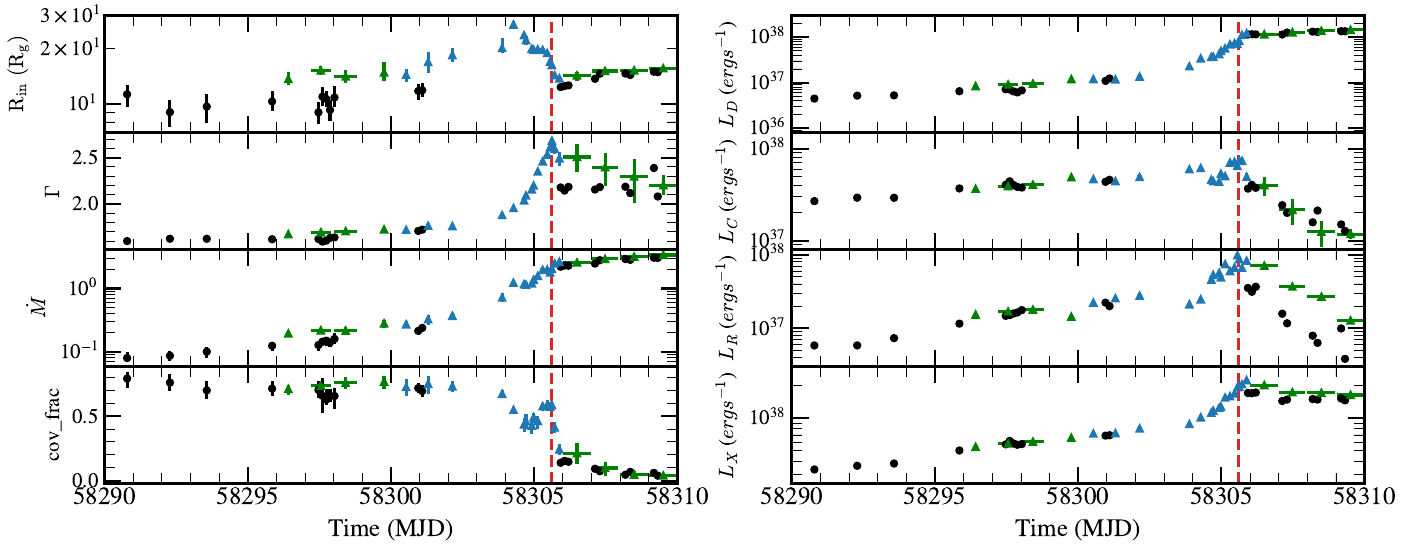}
\caption{The spectral parameters and X-ray luminosities of different spectral components during the IMS when  $Insight$-HXMT observations are absent. The black dots are derived from $Insight$-HXMT observations. The blue triangles are derived from $NICER$ observations we used in \autoref{sec:4.4} and the green triangles are derived from $NICER$ observations simultaneous to $Insight$-HXMT observations. The red dashed line marks the transient jet ejection time and also the boundary between HIMS and SIMS. }
\label{fig:checkgap}
\end{figure*}

We conducted a comparative analysis of the simultaneous $NICER$ and $Insight$-HXMT observations during the IMS when $Insight$-HXMT observations are absent. The evolution of the main parameters and luminosities are present in \autoref{fig:checkgap}. In these simultaneous observations, they show consistent evolutions, although the absolute values of few specific spectral fitting differ. The major discrepancies are $\Gamma$ and $L_\mathrm{R}$ during the SIMS where the values derived from $NICER$ are markedly higher than those from $Insight$-HXMT. The disk component becomes dominated during the SIMS (see \autoref{fig:sedpds}(f)) . So the photon index derived from energy band below 10 keV (similar to $NICER$ data) apparently higher than that from the entire energy band 2--150 keV.

In the HIMS, the Comptonization emission dominates the spectra (see \autoref{fig:sedpds} (d) and (e)), thereby the fitting results are roughly consistent between two telescopes. Therefore, we believe that the fitting results from $NICER$ can indeed serve as a supplement when $Insight$-HXMT obervations are absent during the HIMS.

\section{Some extra materials}
Some additional figures and tables are listed here.

 \begin{figure}
\centering
\includegraphics[width=0.5\textwidth]{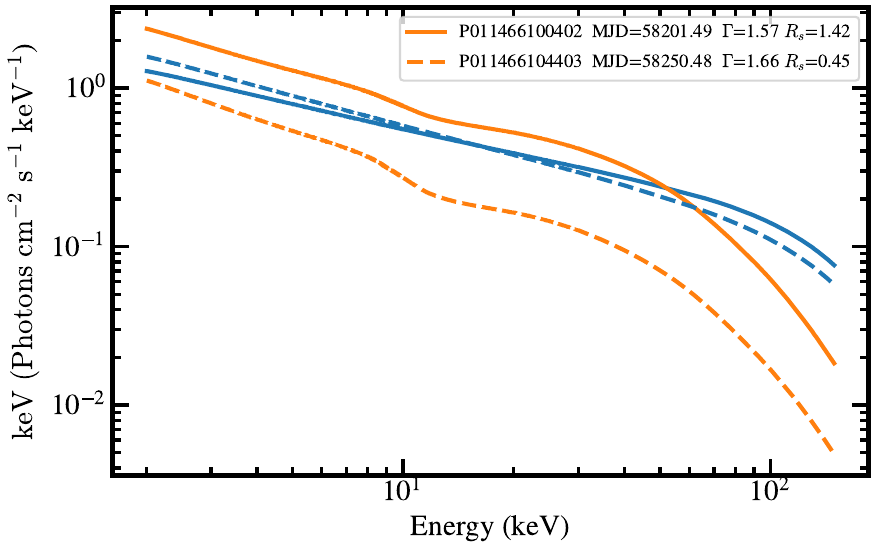}
\caption{The corresponding Compton (in blue) and relativistic reflection (in orange) components of the best-fitting models under different reflection strength $R_s$ in $Insight$-HXMT observations. The exposure ID, $R_s$ and $\Gamma$ are present on the upper right in the figure. It is clear that the Reflection component dominates over the Compton component when the $R_s$ is larger than 1.}
\label{fig:modelref}
\end{figure}

 \begin{figure}
\centering
\includegraphics[width=0.5\textwidth]{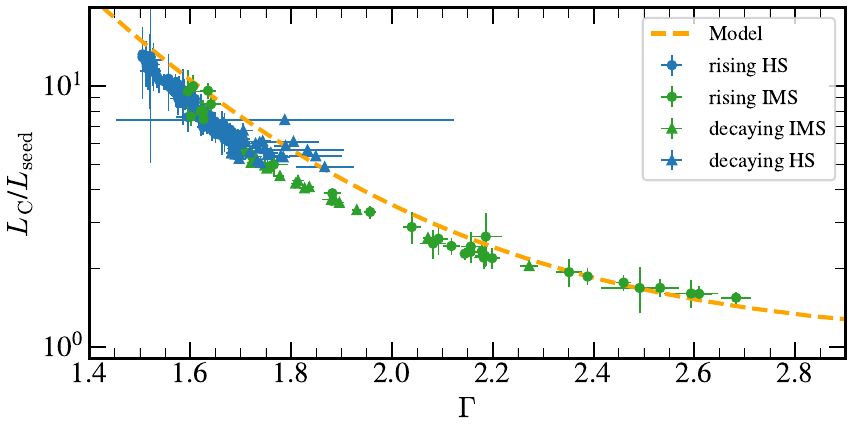}
\caption{The relationship between $\Gamma$ and the ratio of $L\mathrm{_{C}}$ to $L\mathrm{_{seed}}$ is consistent with the model from \citet{Beloborodov1999}.}
\label{fig:LcLseedgamma}
\end{figure}

\begin{table*}
\begin{center}
\caption{The QPO information of each segment in $NICER$ obsID 1200120196 and 1200120197}
\begin{tabular}{ccccccccccc}
\hline\noalign{\smallskip} 
 \multicolumn{1}{c}{$\mathrm{Start}$} & \multicolumn{1}{c}{$\mathrm{End}$} & \multicolumn{1}{c}{Time} & \multicolumn{1}{c}{$\nu_{\mathrm{C,QPO}}$} & \multicolumn{1}{c}{RMS$_{\mathrm{C,QPO}}$} & \multicolumn{1}{c}{Q$_{\mathrm{C,QPO}}$}   &  \multicolumn{1}{c}{$\sigma_{\mathrm{C,QPO}}$}  & \multicolumn{1}{c}{$\nu_{\mathrm{B,QPO}}$} & \multicolumn{1}{c}{RMS$_{\mathrm{B,QPO}}$} & \multicolumn{1}{c}{Q$_{\mathrm{B,QPO}}$}   &  \multicolumn{1}{c}{$\sigma_{\mathrm{B,QPO}}$}   \\ 
  \multicolumn{1}{c}{(s)} & \multicolumn{1}{c}{(s)} & \multicolumn{1}{c}{(MJD)} & \multicolumn{1}{c}{(Hz)} &   &    &    & \multicolumn{1}{c}{(Hz)} &    &    &     \\ 
\hline\noalign{\smallskip}
142237907.0   & 142238873.0 & 58304.2728 & $2.1063^{+0.0197}_{-0.0180}$  & $0.0366^{+0.0017}_{-0.0018}$ & 5.6347 & 20.7906 & -- --                        & -- --                        & -- --  & -- --   \\\noalign{\smallskip}
142269824.0   & 142271003.0 & 58304.6423 & $2.9366^{+0.0201}_{-0.0208}$ & $0.0277^{+0.0011}_{-0.0011}$ & 7.2508 & 24.6821 & -- --                        & -- --                        & -- --  & -- --   \\\noalign{\smallskip}
142275663.0   & 142276722.0 & 58304.7098 & $3.2084^{+0.0194}_{-0.0198}$ & $0.0288^{+0.0008}_{-0.0011}$ & 6.0187 & 30.1199 & -- --                        & -- --                        & -- --  & -- --   \\\noalign{\smallskip}
142292969.0   & 142294468.0 & 58304.9101 & $3.8371^{+0.0361}_{-0.0368}$ & $0.0217^{+0.0012}_{-0.0012}$ & 6.0500   & 17.7883 & -- --                        & -- --                        & -- --  & -- --   \\\noalign{\smallskip}
142298630.0   & 142300026.0 & 58304.9757 & $4.2509^{+0.0225}_{-0.0268}$ & $0.0206^{+0.0008}_{-0.0008}$ & 8.5140  & 25.8217 & -- --                        & -- --                        & -- --  & -- --   \\\noalign{\smallskip}
142308700.0   & 142314902.0 & 58305.0922 & $4.9658^{+0.0324}_{-0.0344}$ & $0.0161^{+0.0009}_{-0.0009}$ & 7.7465 & 18.1335 & -- --                        & -- --                        & -- --  & -- --   \\\noalign{\smallskip}
142319818.0   & 142333386.0 & 58305.2209 & $5.9361^{+0.0284}_{-0.0248}$ & $0.0138^{+0.0004}_{-0.0004}$ & 4.9696 & 37.6554 & $3.9725^{+0.1147}_{-0.1200}$   & $0.0086^{+0.0004}_{-0.0003}$ & 2.7874 & 25.1214 \\\noalign{\smallskip}
142336496.0   & 142344002.0 & 58305.4139 & $6.514^{+0.0347}_{-0.0367}$  & $0.0099^{+0.0003}_{-0.0003}$ & 7.3527 & 29.2912 & $4.4448^{+0.1583}_{-0.1605}$ & $0.0078^{+0.0004}_{-0.0004}$ & 3.2001 & 20.7388 \\\noalign{\smallskip}
142347614.0   & 142350064.0 & 58305.5426 & $7.1084^{+0.0953}_{-0.0904}$ & $0.0106^{+0.0007}_{-0.0007}$ & 3.3794 & 15.9888 & $4.7946^{+0.1202}_{-0.1442}$ & $0.0070^{+0.0005}_{-0.0005}$  & 5.5420  & 14.4097 \\\noalign{\smallskip}
142353173.0   & 142355624.0 & 58305.6069 & $8.0492^{+0.1531}_{-0.1288}$ & $0.0051^{+0.0007}_{-0.0007}$ & 4.4178 & 7.2738  & -- --                        & -- --                        & -- --  & -- --   \\\noalign{\smallskip}
142358915.0   & 142366743.0 & 58305.6734 & -- --                        & -- --                        & -- --  & -- --   & $4.1864^{+0.0275}_{-0.0275}$ & $0.0084^{+0.0003}_{-0.0004}$ & 5.6659 & 22.8935 \\\noalign{\smallskip}
142370610.0   & 142383043.0 & 58305.8088 & -- --                        & -- --                        & -- --  & -- --   & -- --                        & -- --                        & -- --  & -- --  \\
\hline\noalign{\smallskip} 
\end{tabular}
\end{center}
\label{Tab:pds196197}
\begin{flushleft}
Notes:\\ 
$\mathrm{Start}$: the start time of segments; $\mathrm{End}$: the end time of segments; $\nu$: centroid frequency; RMS: root mean square; Q: quality factor; $\sigma$: significance. The subscript B and C represent the type-B and type-C QPOs.
\end{flushleft}
\end{table*}

\begin{figure*}
\centering
\includegraphics[width=1\textwidth]{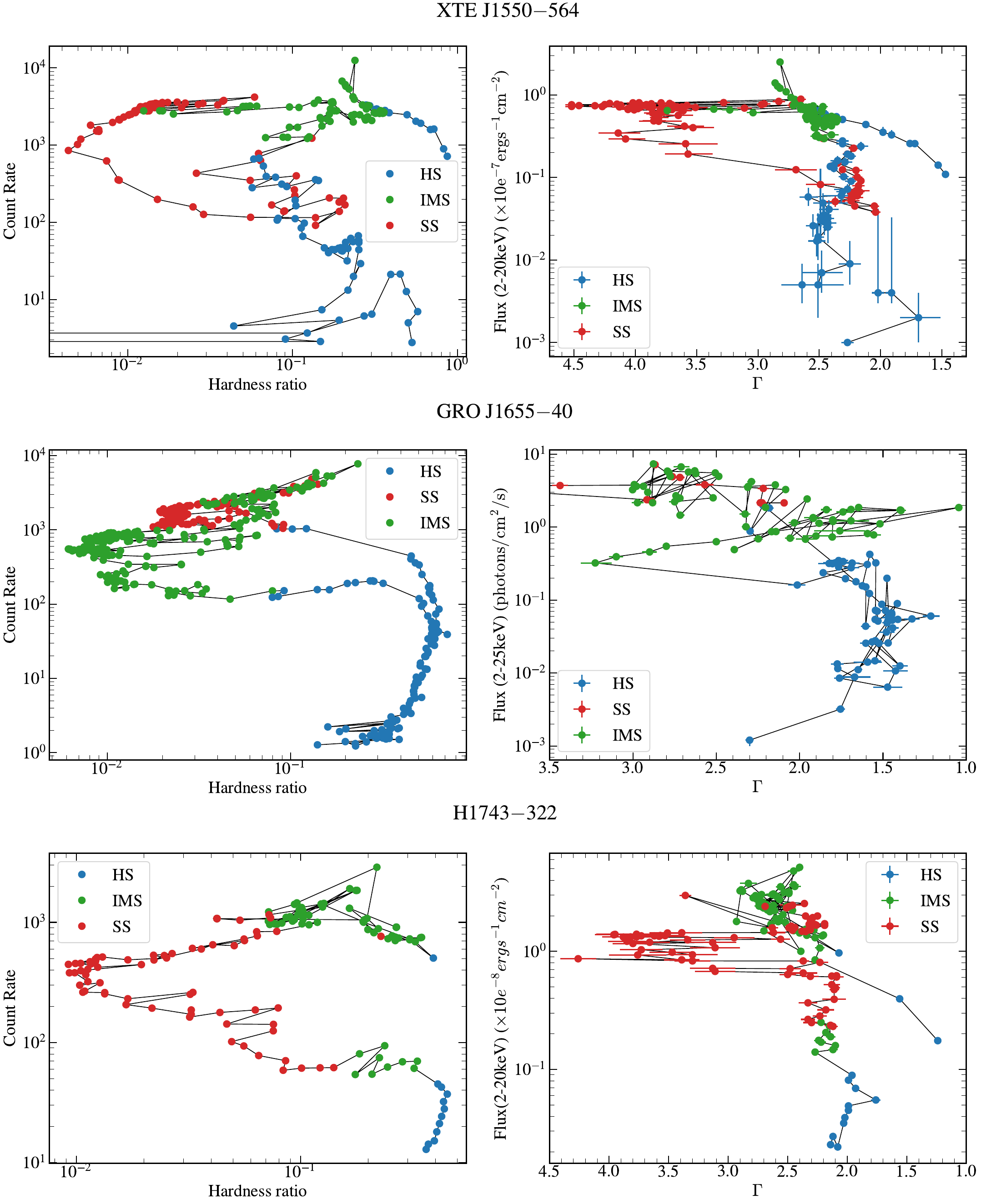}
\caption{The hardness-intensity diagram (HID) and flux-$\Gamma$ plane during three outbursts in other sources, with the left panels showing the HID and the right panels showing the flux-$\Gamma$ plane. The outbursts, in order from top to bottom, are as follows: the 1998 outburst of XTE J1550$-$564, the 2005 outburst of GRO J1655$-$40 and the 2003 outburst of H1743$-$322. The HIDs are defined as the total 2.5--30 keV count rate vs. the ratio of hard (10--30 keV) to soft (2.5--6 keV) count rates from $RXTE/PCA$. The flux and $\Gamma$ data are from public data in \citet{Sobczak2000,Debnath2008,McClintock2009}, respectively.}
\label{fig:others}
\end{figure*}

\begin{figure*}
\centering
\includegraphics[width=1\textwidth]{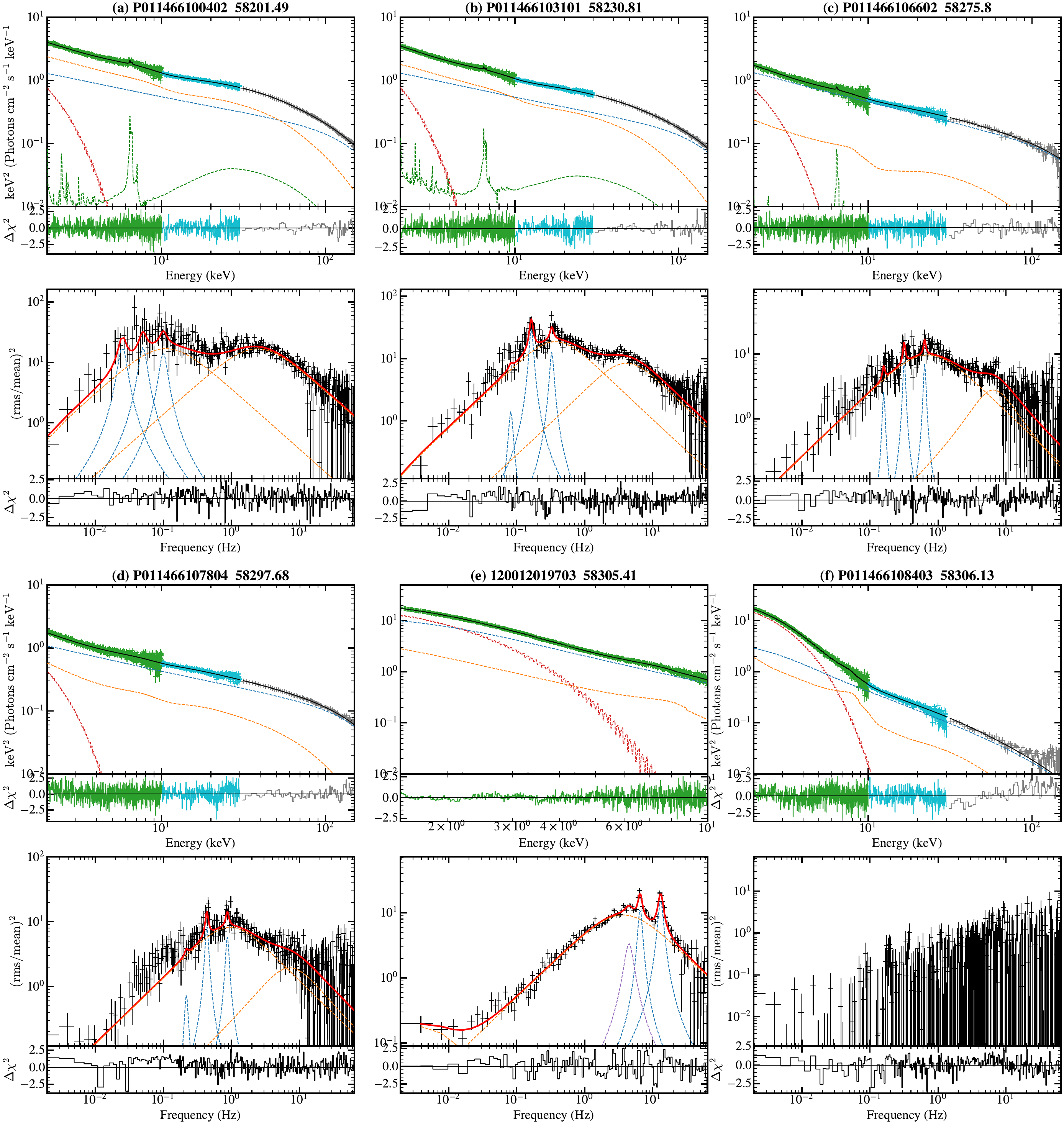}
\caption{Examples of spectra and PDS for five $Insight$-HXMT observations and one $NICER$ observation. These observations capture distinct phases of the HS and IMS: the rise, plateau, and bright decline during the HS and HIMS as observed by Insight-HXMT, the HIMS observed by NICER, and the SIMS. In spectra, the error bars show the data from $Insight$-HXMT LE(green), ME(cyan), HE(black) and NICER(green). The dashed lines show the best-fitting components: disk(red), Compton(blue), relativistic reflection(orange) and non-relativistic reflection(green) components respectively. In PDS, The dashed lines show the best-fitting components: Type-C QPO and its harmonic and/or subharmonic(blue), Type-B QPO(purple) and BLN components(orange).}
\label{fig:sedpds}
\end{figure*}



\bsp	
\label{lastpage}
\end{document}